\documentclass[12pt,preprint]{aastex}
\begin{document}
\tighten
\title{Fluid Dynamics of Stellar Jets in Real Time: Third Epoch HST Images of HH 1, HH 34, and HH 47 }

\author{
	P. Hartigan \altaffilmark{1}, 
	A. Frank \altaffilmark{2}
	J. M. Foster \altaffilmark{3}
	B. H. Wilde \altaffilmark{4}
	M. Douglas \altaffilmark{4}
	P. A. Rosen \altaffilmark{3}
	R. F. Coker \altaffilmark{4}
	B. E. Blue \altaffilmark{5}
	J. F. Hansen \altaffilmark{5}
	}

\vspace{1.0cm}

\altaffiltext{1}{Rice University, Department of Physics and Astronomy,
6100 S. Main, Houston, TX 77521-1892} 

\altaffiltext{2}{University of Rochester, Department of Physics and Astronomy,
Rochester, NY 14627-0171} 

\altaffiltext{3}{AWE, Aldermaston, Reading Berkshire, RG7 4PR, UK}

\altaffiltext{4}{Los Alamos National Laboratory, Los Alamos, NM 87545}

\altaffiltext{5}{General Atomics, 3550 General Atomics Court, San Diego, CA 92121-1122}

\begin{abstract}

We present new, third-epoch HST H$\alpha$ and [S~II] images of three HH jets
(HH~1\&2, HH~34, and HH~47) and compare these images with those from the 
previous epochs.  The high-spatial resolution, coupled
with a time-series whose cadence is of order both the hydrodynamical
and radiative cooling timescales of the flow allows us to follow
the hydrodynamical/magnetohydrodynamical evolution of an astrophysical
plasma system in which ionization and radiative cooling play significant
roles.  Cooling zones behind the shocks are resolved, so it is possible to identify
which way material flows through a given shock wave.
The images show that heterogeneity is paramount in
these jets, with clumps dominating the morphologies of both bow shocks
and their Mach disks. This clumpiness exists on scales smaller than the jet widths and
determines the behavior of many of the features in the jets.
Evidence also exists for considerable shear 
as jets interact with their surrounding molecular clouds, and
in several cases we observe shock waves as they form and fade where material emerges
from the source and as it proceeds along the beam of the jet.
Fine-structure within two extended bow shocks may result from 
Mach stems that form at the intersection points of oblique shocks within
these clumpy objects.  Taken together, these observations represent 
the most significant foray thus far into the time domain for stellar jets, and
comprise one of the richest data sets in existence for comparing 
the behavior of a complex astrophysical plasma flows with numerical
simulations and laboratory experiments. 


\end{abstract}

\section{Introduction}

A variety of objects with accretion disks drive supersonic jets, including 
compact binaries, AGN nuclei, planetary nebulae, and young stars \citep{cbref,agnref,pnref,ysoref}.
Young stars are particularly good places to study the dynamics of these outflows
because the best images easily resolve the widths of jets, and because
postshock gas cools by radiating emission lines, so images and spectra show where the gas is heated.
The cavities, shells, and turbulence created by stellar jets as they inject kinetic
energy and momentum into the surrounding medium are a major way that 
young stars generate feedback to their environments, so by observing how stellar
jets interact with themselves and their surroundings we can identify the main fluid
dynamical processes that operate in these systems. Such observations
motivate and guide numerical simulations of stellar jets, and also
provide some insight into how to interpret other astrophysical jets that are unresolved spatially.

The most simple and powerful way to study the dynamics within a complex system like a stellar jet
is simply to watch how it varies with time. With most astrophysical nebulae this is
not practical because the evolutionary timescales are too long and distances too great
for multiple epochs of imaging to show substantial motions or structural changes in the
emitting gas. A notable exception to this is the Crab Nebula, where `movies' at optical and
X-ray wavelengths revealed a wealth of dynamical information about the jet and its surrounding
synchrotron nebula \citep{crabref}.

Proper motions of HH~objects (the brightest shocks within jets) were first observed decades ago from
ground-based proper motions \citep[][e.g.]{ch79,hj81}. 
These observations show the overall nature of the bipolar flows, but only the
vastly superior spatial resolution provided by HST makes it possible to clearly resolve
cooling zones and individual shocks in each region.
The nature of the line emission from these objects greatly assists in their interpretation. 
Because shocks in stellar jets typically move into gas that is mostly neutral, Balmer 
line emission occurs primarily in a thin shell excited by collisions at the shock front,
followed by a cooling zone that is bright in forbidden lines of [S~II], [O~I], and [N~II]
\citep{heathcote96}. Cooling zones are often observed to be offset spatially from the shock
front, so by combining narrowband H$\alpha$ and [S~II] emission-line images 
we can identify both the location of the shocks and the direction of the flow through the shock.

Papers based on images from two epochs of HST images exist for the bright, well-resolved
jets and bow shocks in HH~1 \& 2
\citep{hh1pmot}, HH~34 \citep{hh34pmot}, and HH~47 \citep{hh47pmot}. These works focused on measuring
precise bulk velocities throughout these regions, and the observations were ideal for this
purpose because the two epochs in each case were separated by $\sim$ 4 years, 
long enough for the emission line objects to show well-defined proper motions yet short enough
to ensure that structural variations remain negligible within most knots. 
In the decade that passed since the second-epoch observations of these regions,
most features in the emission line images exhibit morphological changes that range from dramatic to subtle.
These variations hold the key to understanding ongoing dynamical processes 
by providing the first venture into the time-domain of jets from young stars.

In what follows we describe new HST images of HH 1 \& 2, HH 34, and HH 47 in H$\alpha$ and [S~II]
(section 2) with an emphasis on quantifying the most obvious 
dynamical changes (section 3). 
We group the various dynamical phenomena we observe into five categories of
common physical processes (section 4), and end with a summary (section 5).
While analyzing these data we found that the complex morphological changes we observe
are often easiest to visualize by examining `movies' interpolated from the
three epochs in a uniform manner, with the caveat that any behavior inferred
from the movies must be confirmed by examining the individual images in each filter.
Examples of these movies are available on the web and these are a helpful supplement to
the figures in this paper. 

\section{Data Acquisition and Reduction} 

We obtained a third epoch of H$\alpha$ (656N) and [S II] (673N) images of HH 1 \& 2, HH~34 and HH~47
with WFPC2 as part of HST program GO-11179.  Table 1 summarizes the current and previous two epochs
of these three jets taken with the same filters on WFPC2 over a span of 14 years as parts of three other GO programs. 
The data acquisition strategy differed between the programs, with some opting for a fewer,
longer exposures to minimize readout noise, while others choosing to obtain many shorter exposures 
to facilitate the removal of cosmic rays and bad pixels. In addition, some of the programs dithered
by fractional pixels between exposures to improve the spatial resolution while others did not. The
fields of view also varied, depending on the pointing constraints for the telescope at the
epoch of the observations, a desire to image different parts of the regions, or the need to
avoid a particular WF chip that had subpar dark current or cosmetic defects.

In our current program, for HH~1 \& 2 we dithered each set of exposures between two positions separated by
2.5 WF pixels, using a total of two exposures per dither point in each filter.  Similarly, for HH~34 we used an
analogous 4-point dither pattern with two exposures in each filter, and for HH~47 a 3-point dither pattern with two 
exposures in each filter.  The observations were obtained when HST was operating on only two gyros, and as
a result the optimal orientations for each region were not always available. This constraint was most problematic
for HH~1 \& 2 because the available orientations left almost no room to fit both HH~1 and HH~2 on the CCDs in a
single exposure. We were able to fit the brightest portion of HH~1, and all of HH~2 onto the CCDs with 
clearances of about 0.5 and 2.0 arcseconds, respectively. However, the leading edge of the faint H$\alpha$-bright
bow shock labeled `B' by \citet{hh1pmot} fell just off the edge of the chip. While this object is probably
more interesting dynamically than the knots in HH~2L located near the bottom
edge of the field, HH~2L proved to be an important check of the procedure
used to align and combine the individual chips into the final composite image, because
the HH~2 field has no stars useful for alignment in its vicinity.  The image of HH~47 was shifted with
respect to previous epochs in order to capture the redshifted counterjet for future proper motion studies,
at the expense of not imaging the faint extended bow shock HH~47D. 
The summed images H$\alpha$ + [S~II] appear in Fig.~1 for each of our three fields. 

The optimal procedure for reducing WFPC2 images has evolved over the years since HST was launched.
More is known now about distortion corrections than in the past, and better routines are available to correct for
cosmic rays and bad pixels.  While reduced and aligned images of these regions exist as part of the previous programs,
to facilitate comparing the old images with the new ones we decided to reanalyze all of the previous images so as
to reduce all the data in the same manner. 

A typical exposure time was over 1000 seconds for each frame, so the images contain many cosmic rays. 
In addition, with only a handful of images available for each region, standard procedures for removing
cosmic rays are limited.  In each region we began by removing cosmic rays from the dark-corrected,
flattened images with the Laplacian-based routine la\_cosmic \citep{lacos}.
After performing this procedure for each of the four chips in each exposure, we blinked
the cleaned and unclean images, paying special attention to any area that had extended emission
to ensure that the routine did not remove any real features. Roughly 90\%\ of the cosmic rays were removed 
with this procedure. We then employed the STSDAS/IRAF routine crrej as the next stage for cosmic ray correction,
reviewing each frame interactively as we did for la\_cosmic. 

Images recorded on the four WFPC2 chips need to be corrected for chip-to-chip rotations and distortions, which are
time-dependent. To accomplish this we used a program devised by J.~Anderson at STScI where the corrections
are based in part on astrometric observations of $\omega$ Cen \citep{anderson03}. The dithered images were shifted,
corrected for differing zero-point levels, and combined so that the final composites
have a plate scale of 0.05 arcsecond per pixel. Typical positional scatter 
between different epoch images of the same star are $\sim$ 0.01 $-$ 0.02 arcseconds. For HH~2 we rely completely
on the distortion corrections because no stars are present to check the alignment. One check on the alignment is
to follow the position of the knots associated with HH~2L, which is sensitive to rotational offsets because
it has a low proper motion and is located at the edge of the frame.  Proper motions measured from the new
images used the code described in \citet{hartigan01}.

Ratios of emission-line images can provide important information about physical conditions in extended nebulae.
For HH nebulae there are no significant contaminant lines within the 656N and 673N bandpasses, so 
the counts recorded in these images are proportional to the fluxes in H$\alpha$ and
[S~II] 6716+6731, respectively. The proportionality constant between the count rate and the flux depends on
the level of continuum light present, and on the filter transmission curves \citep{odell99}. Based on 
spectroscopic data of nebulae imaged through these filters, \citet{odell09} determined that the effective
transmission coefficient for the 673N filter was a factor of 1.14 higher than that of the 656N filter.
We have applied this correction to our emission line ratio images, but it is of little consequence because
the variations in the line ratios are typically a factor of two or larger.

Contaminating continuum light is potentially more problematic, because the bandpass of the 673N filter is
a factor of 2.14 larger than that of 656N.  However, aside from the well-known reflected light cavities
around the sources of HH~34 and HH~47, the only source of continuum in our objects is the 2-photon
continuum which is very weak at these wavelengths. For example, using the measurements of \citet{hartigan99} for
HH~47, taking into account the larger bandpass of 673N relative to 656N, the continuum alters the
observed emission line ratio derived from the images by less than 0.4\%, and can be ignored. 
Because the wavelengths of the H$\alpha$ and [S~II] lines are similar, differential reddening is also
negligible.

\section{Results} 

When combined with the previous two epochs, the new HST images make an
ideal data set to study fluid dynamical processes in stellar jets and working surfaces.
Proper motions were already measured with high precision from translational
motions between the first and second epochs in three published papers 
\citep{hh1pmot,hh34pmot,hh47pmot}. The first two epochs are best-suited for measuring
proper motions because the time interval between epochs is long enough for the objects to
show significant movement, but short enough that the morphology of individual features
typically remains constant.  The main power of the third epoch is to
separate dynamical changes from bulk motion everywhere
within each system. In what follows we focus on these structural changes, but also
report proper motion measurements for new knots where appropriate.

\subsection{HH 1 \& HH 2} 

The HH 1 \& 2 are two of the brightest HH objects in the sky, and were among the
first to be studied for proper motions. Using images that date back to
1946, \citet{hj81} found large proper motions of 200 $-$ 350 km$\,$s$^{-1}$ directed to the 
northwest for HH~1 and to the southeast for HH~2. These observations
demonstrated that HH objects are manifestations
of collimated bipolar outflows (with the new, smaller distance to Orion of 414~pc from \citet{odistance},
velocities derived from these studies will be reduced by $\sim$ 10\%).  
The driving source of the HH objects appears to be a highly obscured source known as VLA~1,
visible only in radio continuum \citep{pravdo86}.
Radial velocities are low, indicating the flow lies nearly in the plane of the sky,
and large emission line widths within the bright knots provided some of the first
concrete evidence for bow shocks \citep{hr84,hrh87}.
\citet{hj81} noted that time-dependent variations occurred, including
the emergence of a bright new knot in HH~2 \citep[see also][]{eisloffel94b},
but it was not until HST was launched that it became possible to resolve
and follow individual shock waves and knots well within this region. 

\citet{hh1pmot} combined first-epoch HST images from 1994 \citep{hh1epoch1} with a second epoch
from 1997 to measure the proper motions of over 150 individual
objects within HH~1, HH~2, and the jets in the system.  
In HH~1, proper motions revealed what appeared to be shear, with the apex of
the brightest bow shock apparently pulling away from material along its eastern edge. 
The images clearly resolved the cooling zone structure of the bow shock, with
H$\alpha$ emission located near the shock fronts followed by [S II] emission. 
\citet{hh1pmot} comment that HH 1 appears to be displaced east of the axis of the jet,
and that is the side of the flow is where the wing of the bow shock exhibits shear. 
The enhanced emission along that wing of the bow shock can be caused by an increase
in the preshock density towards the east \citep{henney96}, an effect that would
also distort the shape of the bow shock relative to its direction of motion like that
observed in HH~1. However, the multiple bow shocks
and the offset of the jet's axis from the brightest bow in this object
must arise from other causes, such as precession or velocity variability in the flow.

Two jets are visible near the driving source, VLA~1.
The HH~1 jet becomes visible at optical wavelengths $\sim$ 6 arcseconds from 
VLA~1 \citep[][; Fig.~1]{pravdo86,reipurth93}, though the jet can be traced closer to its source at
infrared wavelengths \citep{reipurth00,garcialopez08}. In their first-epoch images, \citet{hh1epoch1} discovered
that a smaller and fainter second jet, HH~501, emerges from nearly the same location
as HH~1 does but at an angle of about 10 degrees clockwise from the main jet. \citet{hh1pmot} found no evidence for
any interaction between the two jets, and concluded based on proper motions that both originate near
VLA~1, which itself may be a multiple source.  Radial velocities of the H$_2$ emission show both a redshifted
and a blueshifted component within a few arcseconds of VLA~1, suggestive of two sources
\citep{garcialopez08}, although interpretation of these observations is made more difficult
by the orientation of the flow near to the plane of the sky.
Images from the first two epochs show the HH~1 jet to be bright in
[S~II] and much weaker in H$\alpha$, typical for stellar jets \citep{reipurth00}.

HH~2 is one of the most fragmented shocked flows known, with HST images showing dozens of
overlapping clumps and complex morphologies to the line emission \citep{hh1epoch1}. Proper
motions derived from the first two HST epochs provided intriguing evidence for both forward
and reverse bow shocks, time variability, and differential motions 
\citep{hh1pmot}.  The morphology in HH~2 differs markedly from that of HH~1, but the highest
proper motions in HH~2 lie along an axis that passes through HH~1 and VLA~1.

\subsubsection{The HH 1 Bow Shock} 

Our third-epoch images of HH~1 \& HH~2 provide a definitive test of
the conclusions of \citet{hh1pmot}, and also reveal new phenomena.
Fig.~2 summarizes the primary dynamical behavior in the bow shock of HH~1.
The leading edge of the bow shock is green (H$\alpha$), and
is followed by red ([S~II]) or yellow (both H$\alpha$ and [S~II])
in the cooling zone, consistent with the expected structure of a spatially-resolved
shock.  The angled scale bar labeled 'Shear' in the Figure is of fixed length, with its base
located at the intersection point between the left wing of the bow shock and the
clumpy material to the side for each of the images. In 1994 (panel a), the bow shock is smaller than the scale bar, 
but increases in size by 1997 to the length of the bar (panel b), and becomes significantly larger
than the bar in 2007 (panel c). Hence, the third-epoch image confirms that the bow shock 
pulls ahead of the material in its eastern wake, and thereby creates a region of large shear.

The third epoch resolves remarkable new structures in the HH~1 bow shock, as
shown in panels d, e, and f in Fig.~2. In the latest epoch, the left side of the bow shock
has changed from its usual curved shape into a nearly flat surface, and a new `bubble' has
emerged right at the apex. Trailing the flat edge of the bow shock on the left side, we 
find two small knots that have appeared in the third epoch. These new features are more
prominent in H$\alpha$ than they are in [S II]. We discuss these developments in the context of deflection
shocks in section 4.1.

Fig.~2 shows three features that varied significantly in brightness, especially in H$\alpha$, over the
time period of the observations. All are located along the axis of the fast part of the flow.
The top and bottom of the three, named `29' and `L', respectively, by \citet{hh1pmot} and noted
as variable in that paper, continued to fade in brightness into the third epoch and have nearly disappeared
while becoming more diffuse. A new, arc-shaped H$\alpha$ feature has now formed between the two fading knots
at the position of an ill-defined faint knot present in the first two epochs. These features likely
arise from small fast clumps in the jet that overtake slower material and produce
transient shocks (section 4.5).

The right side of the bright bow shock has also developed intriguing new structures in the third epoch. 
That area, labeled `froth' in panel f, appeared clumpy in the first two epochs, but the third epoch 
exhibits even more fine structure, particularly in H$\alpha$ where several sharp arcs have appeared. Some
of these arcs are reverse-facing, as if this wing of the bow shock is overtaking small slower-moving clumps
in its path. One can also imagine the structure in the froth as arising from irregularities in the shape of
the wing of the bow shock as it sweeps up material.

\subsubsection{The HH 1 Jet} 

Between 1997 and 2007 a very bright new feature, labeled H$_J$ in Fig.~3, emerged at the base of the
optical part of the HH~1 jet. This knot is likely to be the same as the [Fe II] knot H of \citet{reipurth00}. In their 
observations from 1998.15 UT, this knot was 4.8 arcseconds from the VLA source. It is currently located
6.9 arcseconds from VLA~1, and so would need a proper motion of 0.22 arcseconds per year to be the same knot.
For a distance of 414~pc, this would imply a velocity of 432 $\pm$ 20 km$\,$s$^{-1}$, as compared with
velocities of 265 $\pm$ 10 km$\,$s$^{-1}$ typical for the other HH~1 jet knots \citep[][; correcting the
460~pc distance assumed in that paper to 414~pc]{hh1pmot}. Taken at face value these numbers imply that
knot H$_J$ has a shock velocity of $\sim$ 170 km$\,$s$^{-1}$, which is a very strong shock for a jet knot.
However, the relative velocity between this knot and the material ahead of it will gradually decline
as the system evolves, on a timescale comparable to the time required for the knot to sweep up its mass. 
The line ratio map in Fig.~4 shows that knot H$_J$ has a higher H$\alpha$ / [S~II] ratio than that
of the rest of the HH~1 jet.

VLA~1 continues to launch knots into the HH~1 jet, and with the aid of the proper motions measured above
we can estimate when they should appear in the optical. \citet{rodriguez00}
cite evidence for an ejection in the radio continuum between 1986.2 and 1992.9 UT.
If one were to identify this event with the bright FeII knots labeled K and L by \citep{reipurth00}, then
the proper motion would be 0.22 $-$ 0.50 arcseconds per year, consistent with that of knot H$_J$.
Taking the proper motions to be the same as we infer for knot H$_J$,
we predict knots K and L of \citet{reipurth00} should make their presence known in the optical between
2013 and 2018 as they emerge from the dusty envelope 6 $-$ 7 arcseconds from VLA~1. These two knots are
both brighter than knot H in [Fe~II], and could be quite impressive when they appear.

Feature H$_J$ has substructure consisting of three smaller knots. 
The knot closest to the source lies along the axis of the jet
and is bright in both H$\alpha$ and in [S~II]. Roughly 100~AU further along in the jet we find
the other two components, one located on either side of the axis of the flow. The lateral edges of these
two knots, but especially the one to the left, are relatively much brighter in
H$\alpha$ (green in Fig.~3; reddish-purple in the ratio map of Fig.~4). These locations likely denote
locations where the jet encounters and accelerates slower material at its periphery (section 4.1).
There is also some evidence from the ratio map that other knots in the HH~1 jet have enhanced
H$\alpha$ emission of the left side.

\citet{hh1pmot} remarked that except for features which emerge from the source (e.g. G1$_J$ in 1997), 
other knots in the jet (E$_J$, F1$_J$, G3$_J$) faded between 1994 and 1997, and these trends generally
continue into the third epoch. Using apertures between 3 and 7 pixels for the [S~II] images we find that
knots E$_J$, F1$_J$ and G3$_J$ faded by $\sim$ 15\%, 15\% , and 30\%, respectively between epochs 1 and 2.
In epoch 3, knot E$_J$ faded by an additional 60\% , and
now has an indistinct form with a surface brightness only roughly equal to that of the background.
Knot F1$_J$ is also now $\sim$ 40\%\ fainter than it was in epoch 2. 
However, knot G3$_J$ recovered about 20\%\ of its brightness and is now only slightly fainter
than it was in the first epoch. This knot has also expanded, and has begun to develop a distinct
bow shock shape. It retains H$\alpha$ emission in its bow, an indication that the shock wave is currently active.
Based on these numbers, a good order of magnitude estimate for the time required
for [S~II] knots to fade by a factor of two appears to be about 10 years when the knots no longer
have active shocks that supply a significant amount of hot gas to their cooling zones.

\subsubsection{The HH 501 Jet} 

Previous observations showed that HH~501 consists of two knots, denoted as A and B by \citet{hh1pmot}.
The leading knot, A, is very bright in H$\alpha$, with H$\alpha$ / [S~II] $\sim$ 4 in the
first two epochs \citep[][; Figs. 3 and 4]{reipurth00}.
Although these knots are aligned with an embedded object known as the X-nebula, the proper motions indicated
that the driving source was closer to VLA~1, and may arise from a companion to that object.
Our new images show that a third knot, labeled `C' in Fig.~3, has appeared in the wake of HH~501B.
Knot A continues to be bright in H$\alpha$, though the flux ratio with the [S~II] lines
is now closer to unity (Fig.~4).  Knot A has also developed a curved shape to its western side.

Our third epoch provides a much larger time interval over which to measure proper
motions, but the downside is that morphological changes like we observe in knot A are more common,
and these influence proper motion measurements.  Morphological changes are less marked in [S~II],
so we used these images and the method described by \citet{hartigan01}
to redetermine proper motions between 1994.6 UT and 2007.6 UT for HH~501A,
HH~501B, and knot G3$_J$ in the HH~1 jet.
Proper motions for knots A, B, and G3$_J$ between 1994.6 and 2007.6
were, respectively, 191 km$\,$s$^{-1}$, 165 km$\,$s$^{-1}$,
and 284 km$\,$s$^{-1}$, at 316.0, 319.0, and 323.0 degrees. Within the uncertainties introduced by
morphological changes, these results agree with those
from the first two epochs of \citet{hh1pmot}, which, when corrected for the different assumed distance
are, respectively, 203 km$\,$s$^{-1}$, 155 km$\,$s$^{-1}$, and 266 km$\,$s$^{-1}$
at 320.0, 319.9 and 328.3 degrees. The position angles from A, B, and G3$_J$ to VLA~1 in the first epoch were,
respectively, 317.3, 318.5, and 324.9 degrees, so the new proper motion vectors for the three knots are
aligned even closer with VLA~1 than the old ones were, within 2 degrees ($\sim$ 0.25 arcseconds)
of the source. Hence, the new data confirm that both the HH~1 and HH~501 jets originate from VLA~1. 

%

\subsubsection{HH 2} 

Fig. 5 outlines the nomenclature associated with HH~2, and Figs.~6 and 7
summarize the morphological changes within the complex. In the northern
half of HH~2 (Fig.~6), knot K roughly doubled in size
between epochs 2 and 3, and is now particularly bright in H$\alpha$. As noted by \citet{hh1pmot}, this object has a 
relatively low proper motion and is a reverse-facing bow shock. The third epoch image indicates it is  
a dense knot in the process of being entrained by a faster wind, and presents an outstanding opportunity to
follow this process in real time.  Knot G is a clumpy sheet-like structure that exhibits strong shear, with
the southern portion steadily moving ahead of the northern portion. Knot B appears to lag behind the other
knots in the region, and is probably located on the near or far-side of the cavity that contains the 
fastest knots. Between knots B and K there is an H$\alpha$-bright arc with a high velocity that has nearly
reached the position of knot G by the third epoch. Fig.~6 also marks the location of
two new forward-facing bow shocks that appeared along the axis of the fastest parts of the flow in the
last epoch.

The southern half of HH~2 is dominated by the brightest knot, H, and the historically most variable one,
A \citep[e.g.][]{hj81}. Fig.~7 shows that knot A is actually a composite between a fast component and a 
slow component, with the brightest region associated with the slower component. This slower knot faded
by factors of $\sim$ 4 in [S~II] and $\sim$ 15 in H$\alpha$ between 1994 and 2007, while
becoming more diffuse. Knot H is also in a period of dramatic change as a fast knot (Fig.~7) has moved
ahead and now appears nearly coincident with the leading edge of the object. A new arc, bright in
H$\alpha$ has appeared in front of the leading edge of knot H. Major photometric changes have also
occurred within knot H, most notably a brightening of a region located near the intersection
of two bow shock wings that emanate from two of the leading clumps in the flow.
This region increased in brightness by a factor of $\sim$ 7 in H$\alpha$
and a factor of $\sim$ 4 in [S~II] between 1994 and 2007, and this region now has a surface brightness
double that of the brightest area within knot H in 1994 in both H$\alpha$ and [S~II].

Despite the morphological complexity within HH~2, the H$\alpha$/[S~II] line ratio map of the region
in Fig.~8 shows a remarkably uniform gradient in the sense that H$\alpha$ is strongest near 
the leading edge of knot H, and gradually declines to the northwest. The contour map displays the
total flux in both filters, and there is some correlation in the sense that brighter
regions tend to have higher H$\alpha$.  We discuss evidence for an ultraviolet precursor in section 4.5.

\subsection{HH 34} 

HH~34 was one of the first jet/bow shock systems identified \citep{reipurth86}.
The system has a prominent jet that emerges to the south of an embedded infrared source
\citep{reipurth00}, and along this direction there is a large, bright filamentary bow shock HH~34S (Fig.~9).
A much fainter jet and bow shock exist on the north side of the flow \citep{bmr88}, and deep wide-field
images trace a series of faint bow shocks over a length of $\sim$ 3 pc, implying a dynamical
time of at least 10$^4$ years \citep{devine97}.  Ground-based proper motions \citep[e.g.][]{hr92}
show that knots in the jet emerge about 30 degrees from the plane of the sky with velocities
of 200 $-$ 300 km$\,$s$^{-1}$. HST images demonstrate that the jet expands with a full opening angle 
of just less than a degree, and the ejection direction varies with time, giving the jet a 
wiggled shape \citep{hh34pmot}.
Other apparently unrelated embedded young stars, including Re 22, exist in the area \citep[see][for a
complete discussion of these sources]{hh34pmot}.

The jet is strong in [S~II], indicative of shock velocities $\sim$ 30 km$\,$s$^{-1}$ \citep{hmr94},
so the velocity variations that cause the shock fronts are $\sim$ 10\% of the flow speed.
The shock velocity in the bright bow shock HH~34S must be significantly higher, $\sim$ 120 
km$\,$s$^{-1}$, to account for the [O~III] emission observed near the apex of the bow, an observation
that also enables an estimate of the preshock magnetic field of $\sim$ 15 $\mu$G
\citep{morse92}.  Projecting the jet to HH~34S, there is a confined region of lower excitation \citep{hr92},
and Fabry-Perot spectra also show the emission line width there are broader \citep{morse92},
implying this region represents the Mach `disk', where the jet impacts the bow shock.  

The second-epoch HST images \citep[][Table 1]{hh34pmot} are much deeper than those in the first
epoch, and clearly resolved the shock waves in HH~34S and along the jet. 
The images of HH~34S revealed that the bow shock consists of several filamentary
H$\alpha$ sheets which outline the shock fronts. A reverse-facing bow shock, visible in H$\alpha$ and
followed by a [S~II] cooling zone, marks the location of the Mach disk.
The jet is especially bright in [S~II], but H$\alpha$ is also visible in the jet, especially along the
edges where weak bow-shock-like wings trail behind several knots.

\subsection{The HH 34 jet} 

The source of the HH~34 jet is invisible at optical wavelengths, but it illuminates a reflection
nebula at the base of the jet and this nebula is present in all our HST images. Fig.~10. shows
that the reflection nebula retains roughly the same morphology between the three
epochs $-$ point-like, surrounded by a faint inverted V-shape typical of opaque flared disks
\citep[e.g.][]{wood98}. As noted by \citet{hh34pmot}, the nebula is bright in
H$\alpha$, with the observed H$\alpha$ / [S~II] ratio $\sim$ 0.7 $-$ 1.1 for the three epochs despite the fact that the
bandpass of the [S~II] filter is a factor of 2.1 broader than that of the H$\alpha$ filter.
Spectra of active accretion disks typically have very strong H$\alpha$ emission, consistent with
the observations and with a published spectrum of the source \citep{reipurth86}. 
The reflection nebula was about 50\%\ brighter in the third epoch
than it was in the first two epochs, and it will
be interesting to see if this presages the emergence of a new knot.

\citet{hh34pmot} identified several knots close to the source in the 1998.7 UT images, and measured
proper motions for them. Our third epoch images extend this work by identifying a new knot, and by
measuring proper motions for knot A5, which first appeared in the second epoch.
The new knot, labeled A7 in Fig.~10, is located $\sim$ 175~AU from the reflection nebula, and,
like other knots near this
position, has a higher H$\alpha$ / [S~II] ratio compared with knots further downstream.
We discuss the excitation and nature of this feature in section 4.4.

Knot A5 was by far the brightest feature in the jet close to the source in 1998.7, but in the new
images it faded by a factor of $\sim$ 2 in [S~II], and disappeared completely in H$\alpha$.
Between epochs 2 and 3 the knot changed its morphology from nearly
symmetrical to being stretched along the flow direction.
The proper motion of knot A5, measured from the [S~II] images in 1998.7 and 2007.8 UT, is 
87 $\pm$ 4 mas$\,$yr$^{-1}$ at PA 167.4 $\pm$ 1.5 degrees, where the uncertainties are derived from defining 
slightly different boundaries for the knot. An additional systematic uncertainty of 10 mas$\,$yr$^{-1}$
arises from the morphological changes.  If we were to trace knot A5 backward in
time it should appear 0.1 $-$ 0.2 arcseconds from the reflection nebula in the first epoch, but there was
no trace of the knot at that time. Hence, the knot either decelerated during this period, or
initially formed a few tenths of an arcsecond from the source.

Fig.~10 shows that knot A4 expanded over the time period while fading by a factor of two in [S~II].
The location of the knot in the three epochs, 0.68, 1.09, and 2.05 arcseconds, respectively,
from the brightest part of the reflection nebula, implies a proper motion of 103 $\pm$ 5 mas$\,$yr$^{-1}$
between epochs 1 and 2, and 105 $\pm$ 5 mas$\,$yr$^{-1}$ between epochs 2 and 3, in reasonable agreement with
the value of 115 mas$\,$yr$^{-1}$ found by \citet{hh34pmot}.  These proper motions are $\sim$ 20\%
higher than values for knots in the jet further downstream.
The FWHM of the knot in the direction perpendicular to the axis of the jet was 0.26, 0.28, and 0.33 arcseconds,
respectively, in the three epochs. Allowing for the instrumental FWHM of the coadded stellar images 
(0.18 arcseconds), the true width of knot A4 was 0.19, 0.21, and 0.30 arcseconds in the three
epochs, implying a full opening angle for this knot over time of 4.6 $\pm$ 1.0 degrees, considerably larger
than the 0.8 degree opening angle of the jet measured by following how the sizes of different knots
grow with distance from the source. The origin of the differing opening angles
 most likely arises from morphological changes in the knot on scales
comparable to the spatial resolution of the images. For example, knot A4 appears more diffuse in the
third epoch than it does in the first two epochs, where it shows a more point-like core, and this
change will increase the measured value of the FWHM of the feature.

The bright part of the HH 34 jet (Figs. 9, 11) exhibits a rich variety of internal structure. As noted
by \citet{hh34pmot}, H$\alpha$ emission tends to be brighter along the edges of the jet
(knots E, F, and G), and near the apices and wings of small bow shocks (knots E, I, and K).
Despite the complexity of this region, the third epoch indicates only modest structural variations occur
here, with most knots simply translating along the jet.
The most notable exception is the eastern
side of knot F, where what appeared as a bend in the jet became nearly pinched off from
the axis of the flow in the third epoch.
This region also became brighter in both filters but especially in H$\alpha$.
Knots I, J, K, and L all faded significantly in [S~II] in the $\sim$ 13 years between the first
epoch and the third epoch (by factors of 1.3, 1.8, 2, and 2, respectively). Surprisingly, both
\citet{morse93} and \citet{nc93} reported [O~III] emission from knot L indicative of a high shock velocity.

\subsection{The HH 34S Bow Shock} 

When projected along a straight line, the jet intersects
the large bow shock HH~34S at knot E$_2$, where a reverse bow shock forms (Figs.~9, 12). 
This `Mach disk' has strong H$\alpha$ emission on the side facing
the source, followed by a fan-shaped cooling zone $\sim$ 1000~AU in length
that emits primarily in [S~II]. Further along the flow direction are
knots D$_1$ and D$_2$, which have H$\alpha$ emission on the side away from the source,
as expected for the bow shock. Hence, the middle panel in Fig.~12 is essentially a
highly-resolved working surface, where the [S~II] emission is sandwiched between
a forward shock (the bow shock; D$_1$ and D$_2$) and a reverse shock (the Mach disk; E$_2$).

As noted by \citet{hh34pmot}, the bow shock consists of about a dozen overlapping filamentary
sheets.  The most dramatic flux variations in the third epoch images of HH~34 occur where these
sheets intersect. For example, in the Mach disk region, knot D$_1$ lies 
near the intersection point of two H$\alpha$ arcs (Fig.~12). This feature brightened by $\sim$
50\% in both H$\alpha$ and [S~II] between 1998.7 UT and 2007.8 UT.
Situated in a complex region with several filaments, the
bright knot B essentially disappeared in the third epoch, replaced by several
fainter diffuse blobs that appear mainly in H$\alpha$.

Knot C is another example of a bright compact feature, which, like knot D$_1$, is 
located immediately behind the intersection point of two H$\alpha$ arcs (Fig.~12).
Similar compact knots near knot B exhibit peculiar proper motions and are 
highly time-variable. \citet{hh34pmot} discovered that the H$\alpha$ arc immediately south of
knot B broke into four compact knots, suggestive of a fluid dynamical instability. Surprisingly,
Fig.~12 shows that these knots did not develop into Rayleigh-Taylor fingers in the third epoch, but
instead retained roughly the same morphology despite having nearly three times longer to
evolve between the second and third epochs as compared with the first and second epochs.  
We discuss these transient features and intersecting bow shocks in terms of Mach stems in
section 4.2.

\subsection{HH 47}

HH 46/47 originates from a low-mass binary system at the core of an isolated globule at the
periphery of the Gum Nebula \citep{dopita78,reipurth00b}. A bright blueshifted jet emerges from
the base of a reflection nebula near the source, and the jet exhibits a wiggled structure until
it encounters a bright bow shock, HH~47A (Fig.~13). HST images revealed a clear bow shock/Mach disk structure
within HH~47A \citep{heathcote96}, and ultraviolet spectra provide evidence
for multiple shocks in this object \citep{hartigan99}.  The redshifted portion of the flow is more obscured,
and is visible at optical wavelengths only as a short faint jet near the source (Fig.~13), and as
a bow shock near the edge of the globule \citep{dopita82, eisloffel94a}. The redshifted flow creates
a large cavity that is outlined well in mid-IR Spitzer images of the region \citep{nc04}.
Deep wide-field images uncovered additional bow shocks
several parsecs from the source \citep{stanke99}. Hence, the observed emission knots come from
velocity variations where material moves into the wakes of previous ejections.

\citet{hh47pmot} measured precise proper motions in the HH~47 jet based on two epochs of HST images,
and found the jet to be inclined 37 degrees to the plane of the sky,
\citep[assuming a distance to the object of 450~pc][]{hh47dist}.
Proper motions near the base of the jet are $\sim$ 270 km$\,$s$^{-1}$, then drop
to $\sim$ 180 km$\,$s$^{-1}$ further out in the jet,
and gradually increase to $\sim$ 300 km$\,$s$^{-1}$ in front of 
bow shock HH~47A. Proper motions within HH~47A are
higher at the apex and lower along the wings, averaging $\sim$ 200 km$\,$s$^{-1}$. The spread of
The jet knots have low-excitation spectra characteristic of shock velocities $\sim$ 34 km$\,$s$^{-1}$
\citep{morse94}, consistent with the observed dispersion of proper motions within the jet.

The first two HST epochs discovered several unexpected dynamical processes
which operate within the HH 47 jet and bow shock. Near the base
of the flow there is a bright, variable, linear H$\alpha$ feature that appears stationary.
\citet{hh47pmot} interpreted this feature as a deflection shock that
occurs as the wiggling jet impacts the edge of the cavity outlined by the reflection nebula. Many of the
knots in the jet have distinct H$\alpha$-emitting tails that represent wakes from the knots, and when the knot
is displaced from the central axis of the jet the wake occurs primarily on the side nearest the
edge of the cavity.  The Mach disk is a variable structure that developed two C-shaped intrusions in the second
epoch which could come from small dense clumps in the jet as they penetrate the working surface, and
variable filamentary structures also exist within the bow shock. Additional evidence for a clumpy
flow comes from a very bright knot within HH 47A that 
has a significantly larger proper motion than that of the rest of the bow shock.

\subsection{ The Jet and Reflection Nebula in HH 47} 

A color montage of the source region for the three epochs (Fig.~14) confirms that the deflection
shock is a stationary structure. Unlike the reflection nebula in its vicinity, the deflection
shock has no [S~II] component. The reflected light that outlines the flow cavity
is strong in H$\alpha$, with H$\alpha$/[S~II] $\sim$ 1.3 throughout the region.  Hence, the source must possess a
strong H$\alpha$ emission line in its spectrum.  Most parts of the reflection nebula
increased in brightness by 10\%\ $-$ 20\%\ between the first two epochs, and then faded
by about a factor of two in the third epoch.

Near the source, the bright jet knot J$_1$ expanded from a compact C-shaped
morphology in 1994 to a larger less-distinct structure in 2008. Closer to the source,
the jet continues to travel to the right of the average jet direction
defined by the position of the large bow shock HH 47A (up in Fig.~14).
The H$\alpha$ / [S~II] line ratio along the jet is $\sim$ 1.1, with a variation of $\sim$ 15\%.

The lower section of the HH~47 jet (Fig.~15) is a highly structured region that contains hundreds
of interconnected knots and wisps, but despite this complexity the three epochs show that most of this
structure simply translates along the axis of the flow in unison.
\citet{hh47pmot} measured proper motions within this region by
following small sections in the H$\alpha$ and [S~II] images, labeled Jhn and Jsn, respectively,
where n is a sequence number. The H$\alpha$ emission is somewhat
more filamentary than [S~II], and tends to lie between the brightest [S~II] knots. The lowest knot
in this part of the flow, Jh3, is unusual in that it has the shape and cooling structure of a reverse
bow shock. The third epoch image in Fig.~15 confirms this idea, as we now observe that the
two small knots in the wings of Jh3 have detached from the object as they flowed past it. 
Time-interpolated movies of this portion of the jet reveal subtle changes along the flow, especially
within the Jh4$-$6 region, where the leftmost (southeast) knot gradually becomes more detached from the
rest of the jet to its right. The H$\alpha$ / [S~II] ratio declines from $\sim$ 0.35 in the Jh4$-$6
region to $\sim$ 0.2 in Js12.

The upper section of the HH~47 jet is relatively fainter than the lower part, and consists of a series
of H$\alpha$ arcs that lie along the edges of [S~II] knots and filaments \citep[][Fig.~13]{heathcote96}. Motions in this 
section of the jet were relatively uniform between the first and second epochs, and structural
variations were minor \citep{hh47pmot}. The main event in the third epoch is the sudden appearance of
a new knot, labeled Jh13 in Fig.~16.  This knot emits in both H$\alpha$ and in [S~II], and has
a curved shape in the sense of a reverse bow shock where the apex of the bow faces toward the source.
However, the knot is still too small, $\sim$ 0.2 arcseconds in 2008, to have a clearly-resolved bow shape.
Fig.~16 also reveals that the H$\alpha$ filament Jh12 changed from a linear feature to one
that appears comma-shaped.

\subsection{The HH 47A Bow Shock and Mach Disk} 

The bright bow shock HH~47A marks the end of the continuous blueshifted
jet in HH~47 (Fig.~13), and displays a classic bow shock/Mach disk structure.
The Mach disk is clearly-defined as a linear H$\alpha$ emission feature located
at the base of the bow shock \citep[][Fig. 17]{heathcote96}. 
The area between the bow shock and the Mach disk is clumpy and
radiates in both H$\alpha$ and in [S~II].
\citet{hh47pmot} measured proper motions for 12 knots in H$\alpha$ and
20 in [S~II] in HH ~47A (labeled Ahn and Asn where n = 1 $-$ 12, and 1 $-$ 20, respectively),
and found that while the motions were overall consistent with a bow shock structure,
there were several anomalies.  In the image from the second epoch (1999.2 UT), 
two small clumps appear to penetrate into the Mach disk, while in the bow shock
the compact bright clump Ah8 pushed ahead of the other features in the bow on its way
to the front of the bow shock. These observations, combined
with the overall bumpy appearance of the bow shock led to the idea that small clumps
within the jet propagate into the working surface and generate the clumpy structure
present in the bow shock.  Complicating this scenario was a region 
close to the Mach disk where the morphological structure was too variable 
to follow proper motions.

The third epoch images provide a crucial diagnostic of the dynamics within HH~47A,
confirming some of the previous ideas and challenging others. Fig.~17 
presents the three epochs in H$\alpha$ and [S~II].
The top panel is a color composite centered on the Mach
disk, which, although morphologically variable, is present in all three epochs. For
reference, a faint star at the upper right 
marked with a circle illustrates the bulk motion
of the bow shock with respect to the background stars. 
As noted by \citet{hh47pmot}, the jet knot labeled A1 moves faster than the rest
of the bow shock and should catch up to the working surface in about 2030. The third
epoch confirms this motion. This knot
lies to the right of the center of the Mach disk in Fig.~17, and is accompanied
by a counterclockwise tilt in the shape the the HH~47A bow shock. The tilt is in
the sense one would expect if the Mach disk were to move to the right in
response to the changing impact point of the jet knots. The dynamics of this region
will be particularly interesting to follow in the future because the region to
the right of the HH~47A Mach disk is a source of
shocked H$_2$ gas, so the medium into which A1 will be moving will have a
molecular component \citep{eisloffel94a,curiel95}. This molecular hydrogen appears
to be moving with the bow shock, as it has a large radial velocity
\citep{eisloffel00,sg03} and a high proper motion \citep{micono98}.

Fig. 17 shows that the dozen or so knots in the bow shock do not move as one might 
expect from simple dense bullets. Instead, the third epoch reveals a bewildering
variety of variations as new features appear while others fade away. For
example, the two small knots noted by \citet{hh47pmot} and marked with small black
arrows in the middle panel of Fig.~17 penetrate into the Mach disk in the second
epoch, but the third epoch shows that these two features do not continue as new
bright clumps into the working surface, but instead expand and fade into a mottled
background. Likewise, knot Ah8, which appeared to be the best example of a dense knot
moving through the working surface with two epochs of data, faded and
split into two pieces in the third epoch. Similarly, knot Ah10 at the leading
edge of the bow, and knot Ah5 near the Mach disk faded considerably and split
in the third epoch. At least three new bright knots have formed in the third
epoch, labeled Ah13, Ah14 and Ah15 in Fig.~17. The circular feature whose left edge
is defined by knot Ah11 also changed its appearance as the area expanded over the
course of the observations. We discuss what these variations imply for the dynamics within
bow shocks and Mach disks in sections 4.2 and 4.3. 

\section{Discussion} 

In this section we group the main dynamical processes evident from
the three epochs of HH 1 \& 2, HH~34, and HH~47 images
into five broad categories: (1) interactions of jets with cavities and obstacles,
(2) structures within bow shocks, (3) structures within Mach disks,
(4) knot formation in jets, and (5) radiative shock physics, including
precursors, cooling zones and time variability. In each section we collect
evidence from the three regions that illustrate these five dynamical phenomena.

\subsection{Deflection Shocks, Cavity Evacuation, and Entrainment} 

While most of the shock waves in stellar jets arise from velocity variations
in the flow, shocks also occur when the jet deflects from an obstacle 
along its path. The most well-known example of this type of interaction is
HH~110, where the jet changes direction and produces a spray of shock-heated gas
\citep{riera03,lopez05}.  New near-IR images of this system show that the jet 
entrains molecular cloud material as it deflects from the cloud, and recent
scaled laboratory experiments have documented how this process evolves with time
\citep{hhlab}.  By observing such dynamical structures as they change 
with time we can gain some insight into how jets supply energy and momentum to maintain
evacuated cavities.

Several lines of evidence indicate that the brightest bow shock in HH 1 is currently
undergoing a glancing collision with a molecular cloud or cavity wall.
The shear on the left wing of the bow shock (Fig.~2; section 3.1) implies a stronger interaction
occurs there than on the other side of the bow shock, and H$_2$ images reveal that shocked
molecular gas lies along the left side of the bow shock \citep{davis00}, similar to the situation
in HH~110. In addition, the axis of symmetry of the bow
shock points to the right of its proper motion, consistent with a shape distortion
brought on by increased drag on the left-hand side \citep{henney96}.

Images from the most recent epoch show that the bow shock has flattened dramatically near the
apex on the left side, the signature of a new strong deflection shock, and the bubble
ahead of the bow shock is another consequence of this interaction. The small knots that
developed in the wake of the deflection shock along the left bow wing may result
from Kelvin-Helmholtz instabilities brought on by the enhanced shear.
If this interpretation is correct, the new knots on the left side of HH~1 should
grow in size and number as they lag behind the flow to form a structure similar to that observed
in the 1994 image (Fig.~2).  The deflection of the HH~1 bow shock is very oblique,
and this type of situation must be a common occurrence as wiggling jets encounter cavity walls. 

The dozen or so clumps that make up the feature labeled froth in the HH~1 bow in Fig.~2 
could simply arise as the bow shock moves through a clumpy preshock medium.
In this case, as for nearly all HH flows, the 
preshock material is slower jet gas. If clumps in the preshock gas move into the bow shock they
should create small wakes shaped like reverse bow shocks. Alternatively,
if the bow shock were to become `bumpy', then wakes should trail behind 
protuberances in the surface of the shock. Clumps in the froth are near the spatial
resolution of the images, but a few show reverse bow shapes, supporting the clumpy
preshock medium scenario.

Our images of HH~47 (Fig.~14) provide clear confirmation of the deflection shock described by \citet{hh47pmot}.
In this case the shock appears as a steady-state feature, created as the jet moves to the right
relative to the cavity outlined by the reflection nebula. The bright linear
H$\alpha$ emission at the location of the shock is oriented in the right direction to deflect the jet. 
The current flow direction should continue to bring the jet into contact with the edge of the cavity
in the upcoming decades, so the deflection shock in HH~47 should remain bright for the foreseeable future.

Entrainment of material within jets occurs in several of our images as fast material overtakes slower
gas. Examples of reverse bow shocks forming around
slower material include knots Jh3 and Jh13 in HH~47 (Figs.~15 and 16, respectively), HH~2K (Fig. 6),
and knot E$_2$ in the HH~34 bow shock (Fig.~12). 
New forward-facing bow shocks also occur when dense knots in the jet sweep up slower material, as
pointed out by the central arrow in Fig.~2c in HH~1, new forward bow shocks in HH~2 (Figs.~6 and 7), and 
the brightening of knot E in HH~34 (Fig.~11).
Entrainment on smaller scales can result from small changes in the jet's orientation.
The observational signature is the presence of a brighter H$\alpha$ wing on the side of the
jet farthest from the main axis of the flow. Examples include knots H$_J$, G1$_J$, G3$_J$ and G4$_J$
in HH~1 (Figs.~3 and 4), knots I, G, and F in HH~34 (Fig.~11), and knot Js12 and Jh12 in HH~47 (Fig.~15).

\subsection{Intersecting Shock Waves, Mach Stems, Clumps and Sheets} 

Most emission-line features within our images translate along the
direction of motion and exhibit only gradual morphological changes.
However, some bright knots located at the intersection
points of overlapping filaments within the complex bow shock HH~34S
exhibit lateral pattern motion as the intersection point moves. The best example in HH~34S
may be the bright knot A (Fig.~12), which moves to the right when compared 
with other features in the bow shock.  In this same region bright knots
come and go along the curved filament below knot B, which itself has nearly
vanished in the third epoch. Knot C in HH~34S is a fine example of a bright knot with a resolved
cooling zone at the intersection point of two filamentary shocks.

Studies of intersecting shock fronts date back to the original work of
\citet{mach}, and a large body of work has been published concerning
Mach reflection phenomenon, a field that studies the shock waves which
develop as a curved shock moves along a surface \citep[e.g.][]{bendor}. 
In the case of two intersecting
bow shocks, the plane of symmetry of the problem acts as the reflecting
surface. Depending upon the specific heat ratio of the gas,
bow shocks may develop a flat Mach stem at the intersection point as 
the angle of incidence between the bow shocks becomes more oblique. Numerical simulations have
not yet quantified how cooling affects Mach stem formation in astrophysical
shocks, but because cooling effectively lowers the adiabatic index of a flow,
and analytical studies show Mach stems form more readily when the adiabatic
index is large \citep{gammaref}, radiative cooling should suppress the onset of Mach stems
to some degree.  Mach stems have now been observed in laboratory experiments of
strong shocks in the plasma regime \citep{foster10}, and new experiments designed
to explore instabilities in radiative 
blast waves show structures that could be Mach stems \citep{edens10}. 

Because multiple filaments appear to be common in large HH bow shocks,
intersections between filaments are also common.  If a Mach stem forms 
at these locations then the incident gas experiences a normal
shock instead of an oblique one, which will lead to higher compressions
and temperatures than would occur if the Mach stem were absent. 
Hence, the main observational consequence of forming a Mach stem will be to
create a hot spot with a high proper motion that may also have its own pattern
motion as the intersection point varies with time.
Even if Mach stems do not form, the intersection points between multiple
bow shocks should produce local density enhancements that exhibit
pattern motion with respect to the rest of the emission in the bow shock.

Intersecting shocks are the most likely explanation for the pattern motions and
transient knots we observe in HH~34S, but the phenomena
may also occur in HH~47A, where what appeared to be a bright knot with
an unusually large proper motion (knot Ah8 in Fig.~17) suddenly faded in the third epoch. 
Morphological complexities within HH~2 may also arise in part from intersecting shocks.
Another explanation for the variable structures within HH~47A is that 
both H$\alpha$ and [S~II] are optically thin, so regions within large bow shocks
will become brighter and fainter as portions of unrelated shocks
become projected along a common line of sight.

Intersecting shock fronts exhibit another behavior in HH~2, where multiple clumps
and knots have become organized into a sheet-like structure that resembles
a single large bow shock (Fig. 7). Over the period of the observations, knot A faded
and knot H at the leading edge of the fast flow now dominates the emission. 
The excitation map of this region now resembles that of a single large bow shock (Fig.~8).
A major outstanding question in this region has always been why HH 1 and HH 2 look so different
even though they represent opposite sides of the same flow, with 
HH~1 dominated by emission from a single well-defined bow shock, and HH~2 looking like
a clumpy chaotic region. This dichotomy appears to be changing, as the slower clumps in HH 2 
fade away and the fast ones merge into a single structure. It may be that
we have simply been observing HH 2 at an unusual time in its history during the last few decades
when the jet fragmented for some reason, perhaps because it ran into a large obstacle.
If HH~2 continues to evolve as it has in the last two decades, it should become dominated by
knots G and H, and show entrained features like knot K along the sides (Fig.~5). Likewise, HH~1 will
be dominated by knots F and E, with entrainment knots located primarily along its northern side.

\subsection{Mach disk structures} 

Mach disks are shock fronts that decelerate jets at working surfaces.  In 3D-numerical
simulations the geometry of these structures is typically much more complex than that of
a simple disk, and often consists of a mixture of several oblique and normal shocks that may
fragment to produce clumpy structures and filaments \citep[e.g.][]{stone94}.
Simulations also show that the Mach disk region can be highly variable in time, and that working
surfaces may collapse into plugs when the gas cools strongly \citep{blondin90}.
Observationally, Mach disks should form a sharp H$\alpha$ feature on the jet-side of the working
surface as long as neutral material from the jet continues to flow into the working surface.

Our H$\alpha$ images show two well-resolved Mach disks, in HH~47A and in HH~34S,
but these two objects exhibit markedly different morphologies.
The Mach disk in HH~47A is a relatively linear feature that extends over the width of the
jet (Fig.~17), as expected from a simple model. In contrast, what passes as the Mach disk in HH~34S is 
actually a compact
reverse bow shock that forms as the jet flows past a narrow impact point (Fig.~12, middle panel). This type of
deflection may help explain why multiple arcs occur in HH~34S as
opposed to the single bow shock in HH~47A.
The large bow shock HH~1 has no clear Mach disk, a situation that could occur if the velocity in the
jet declined enough to allow the bow shock to move ahead of the jet in the flow.

The shapes of the Mach disks in HH~47A and HH~34S vary with time, but not
significantly more than the rest of the shocked gas does. The most dramatic variation occurred in the
HH~47A Mach disk in 1999, when it developed two small irregularities (black arrows in the middle
panel of Fig.~17) that appeared to be dense jet knots that had begun to penetrate into the working surface. However,
in the third epoch these knots faded and expanded, so their nature is uncertain.
Dense knots should have produced distinct bow shocks as they moved into the working surface.
The third epoch confirms that the fast knot A1 in Fig.~17 continues to maintain its integrity
as it approaches the working surface. It will be very interesting to observe how the Mach disk
changes in response to the impact of this knot in the upcoming decades.

\subsection{Formation of New Jet Knots Near Source} 

The ultimate source of energy for stellar jets is accretion through the circumstellar disks, as
evidenced by the observed correlation between accretion and outflow rates \citep{heg95,cabrit07}.
In a simple model one would expect that episodes of high accretion would result in a jet knot, but
this correlation has not been firmly established. It is also possible that energy from disk
accretion is stored magnetically and subsequently released as plasmoids in reconnection events.
Disk wind models typically become kink unstable beyond a few Alfven radii \citep{ouyed97},
a behavior also observed in laboratory experiments of magnetized plasma jets \citep{bellan05}.
One advantage of magnetized plasmoid ejections is they provide a means to remove magnetic flux
from jets, which allows jets to be driven magnetically close to the source but still support
low velocity shocks at larger distances that would be sub-Alfvenic if not for the reconnection
processes \citep{hartigan07,hmr94}.

Hence, there are essentially two ideas, (1) an accretion-driven scenario where
episodic accretion events in the disk immediately eject plasmoids at different
velocities that later become jet knots, and (2) a magnetic model where accretion-driven magnetic disk winds 
load the jet with matter and energy, but magnetic instabilities within the jet create knots with
supersonic velocity differences.  Although both models predict a correlation between mass accretion
and outflow rates, there are observational differences between the two concepts.
In an accretion-driven model, knots emerge directly from source, so there should be a
one-to-one correspondence between accretion events and observable emission features in the jet. 
In contrast, knots in a magnetic model originate where reconnection occurs, some distance from
the source, so a gap should exist between the source and where the knot first appears. In this
case a knot will not necessarily be caused by a particular accretion event, unless that event
also leads to a magnetic reconnection.

The ideal test between the two ideas would be to identify new knots as they emerge using
a sequence of images taken no more than a few months apart for an object where the jet or
jet cavity is visible all the way to the source. A multi-epoch study of the flow from
XZ~Tau binary by \citet{krist08} shows a new H$\alpha$ shock in 2004 about 2 arcseconds (280~AU)
south of the stellar pair that was not present in 2002 (typical proper motions for the flow are
0.2 arcsecond per year).  However, previous epochs show no evidence for a stationary shock at
this location, so the best interpretation appears to be that a shock formed here in
response to the collision of a faster ejection overtaking a slower one. This system differs from a
typical HH jet in that it ejects multiple shell-like bubbles, though denser knots exist along the axis.
The faintness of the bubbles relative to the bright PSF of the binary pair makes this system a
challenging one to study.

With our current data set for the bright HH~34, HH~1 and HH~47 jets we have only three epochs,
but these provide some insight into the nature of jet launching. The best object in our sample to study knots
as they emerge is HH~34 because there one can trace the jet close
to the accretion disk.  As described in section 3.1, new knots also emerged in the HH~1 jet between the different epochs,
but these were known before from VLA and infrared observations, and became visible as they moved beyond
the obscuring shield of the dense circumstellar disk, and for this reason are less useful for
constraining how jets eject knots from their sources.

Fig.~10 shows that an H$\alpha$-bright knot appears $\sim$ 100 AU from the HH~34
source in all three epochs (knots A4, A5, and A7, respectively).
There is no evidence for bubbles or shells like \citet{krist08} found in XZ Tau.
With only three images we cannot state with any certainty
if these knots emerged from the source, or if they formed close to where we observe them in each epoch.
In recent HST slitless observations of another jet, HH~30,
evidence for heating events $\sim$ 50 $-$ 100 AU from the source comes from 
a sudden rise in the ionization fraction at that location \citep{hm07}, and it is possible we
are observing a similar phenomenon in HH~34. However, a dedicated program of
monitoring with adaptive optics or space-based imaging is needed to address this problem thoroughly.

\subsection{Cooling Zone and Precursor Structures, Fading and Brightening of Knots} 

Sharp H$\alpha$ emission features followed by a more diffuse cooling zone of forbidden line emission
are the primary diagnostic of shock waves in stellar jets \citep{heathcote96}, and in section 3 we
discussed several examples of resolved cooling zones in our data. Because H$\alpha$ emission occurs
as preshock neutral gas enters the shock, H$\alpha$ responds immediately
to changes in the preshock density; in contrast, intensities of the forbidden lines represent a spatial average
over the entire cooling zone.  As discussed in section 4.1, the most variable
features in our images are H$\alpha$ arcs, as expected for shocks that encounter
changes in the preshock density. New features typically appear only in H$\alpha$, or in both H$\alpha$
and in [S~II], but not in [S~II] alone. 

In very complex regions that have multiple knots such as HH 2 (Fig.~7), or for structured
bow shocks like HH~47A (Fig.~17), flux variations
can arise as differential motions align physically unrelated optically thin
filaments along a common line of sight, so to learn about cooling zone timescales
we choose to study relatively isolated knots
that exhibit secular variations in their forbidden line brightnesses.
For example, in HH~34 (Figs.~9, 11), the timescale for knots
I, J, K, and L to fade by a factor of two in [S~II] is $\sim$ 15 years, and knot B within the
HH~34 bow shock (Fig.~12) also fades on a timescale of this order.
Similarly, in HH 501 (Figs.~3, 4), the decline of the H$\alpha$/[S~II] ratio in the jet knots
is an indication of the time to develop a cooling zone once a shock forms, also about a decade.

As described by \citet{hrh87}, the time for a steady state shock wave to cool to $10^4$K
depends inversely on the preshock density, and ranges between 0.2 to 43 years for shock velocities
between 40 and 60 km$\,$s$^{-1}$ and preshock densities between 100 and 1000 cm$^{-3}$.
The time dependence of line emission within resolved cooling zones has been recently modeled
with 2D and 3D simulations by \citet{raga09}. For comparison with our data we are most interested in
predictions of rise and decay times of line fluxes in response to velocity perturbations.
Predictions for time-dependent 1D radiative shock models \citep{hr93,massaglia05} indicate
rise and decay times of $\sim$ 10 years, depending on the input parameters, with 
faster timescales associate with more strongly magnetized flows.
The observed cooling times of about a decade agree well with predictions from these models.

When shock velocities become large enough 
\citep[$\gtrsim$ 90 km$\,$s$^{-1}$ for planar shocks][]{cr85},
the postshock gas emits enough UV radiation as it cools to
ionize incoming hydrogen atoms. This radiative precursor produces a miniature H~II region ahead of the
shock, and when shock velocities are large \citep[$\gtrsim$ 200 km$\,$s$^{-1}$][]{ds95}
emission from the precursor contributes significantly to the observed spectrum.
Radiative precursors have also been observed in laboratory experiments of strong shock waves
\citep{remington06,bouquet04}.  Shock velocities within HH~2H must be
$\sim$ 160 km$\,$s$^{-1}$ to explain the observed emission line widths and the high-excitation
lines of O~III and C~IV \citep{hrh87}. Models of the bow shocks in this region that take the ionizing  
radiation into account improved the agreement with the observed line ratios over previous
efforts \citep{rhh88}.  With a shock velocity $\sim$ 160 km$\,$s$^{-1}$ inferred from these models,
HH~2H should produce enough radiation for the precursor to
ionize preshock gas, but postshock line emission will still dominate the spectrum.

Our images support the notion that radiation affects the strongest shocks within HH~2.
Fig.~8 shows there is an area of diffuse emission $\sim$ 150~AU wide that
follows along the bright leading edge of knot H. This area has a high H$\alpha$/[S~II] line ratio
($\sim$ 5$-$10) typical of photoionized gas. \citet{ds95} comment that the hard ionizing spectrum
from a high-velocity shock produces a large region of partially-ionized hydrogen ahead of it, 
and the preionization at the apex of a bow shock will also be less than that of a planar
shock of the same velocity. Both effects tend to help neutral hydrogen survive the journey to
the shock front, and any neutral H that does make it to the front will produce strong
Balmer emission there, followed by a cooling zone that radiates both forbidden lines
and additional Balmer emission.  The overall structure we observe in Fig.~8 is consistent
with this picture. A strong jump in the H$\alpha$ flux marks the location of the shock, 
preceded by a diffuse region of H$\alpha$, and followed by both strong [S~II] and H$\alpha$
emission.

\section{Summary}

In this paper we have presented high-resolution, multi-epoch HST observations of
three protostellar jets.  Using the epochs as ``frames in a movie", we are able to
articulate the evolving plasma dynamics of these systems with a clarity that has
not been possible to achieve in previous studies.  Our results provide a rich
data set which can be mined by those seeking to understand jet systems across
the spectrum of astrophysics as well as a those studying astrophysical plasmas
in other environments where complex hypersonic flows are present
(SNe, Planetary Nebula, LBVs and AGN). 

Before we summarize our specific results for young stars, it is useful
to consider the broader astrophysical context of time-dependent imagery as a tool for
understanding how plasmas behave.
The ability to accurately model environments as diverse as cosmological galaxy
formation and space weather in the form of coronal mass ejections interacting
with the Earth's magnetosphere all rely on understanding the complex, time-dependent
evolution of plasma dynamical systems.  Because the governing
equations for these systems are multi-dimensional, coupled, non-linear partial
differential equations, theoretical exploration via analytic methods can only
account for the simplest cases.  Even when numerical simulations are brought
to bear, the inherent multi-dimensionality of the problems of interest,
particularly when magnetic fields are involved, often makes it impossible 
for the simulations to resolve all the important features of the flows.
This limitation can be particularly vexing when
the flows are inherently multi-scale, as is the case for isotropic turbulence and for
inherent heterogeneity (clumpiness). 

Additional complications arise when the
problem involves ``multi-physics", i.e., a number of physical processes at work
simultaneously such as self-gravity or strong cooling by optically thin
radiative line emission.  In these cases a range of time and length scales exist
for the problem which must each be resolved computationally.  This can, for
example, occur in cooling flows in which one has a hydrodynamic times scale $t_h$
$\sim$ L/c (L is the size of interest, c is the sound speed), a cooling
timescale $t_c$ $\sim$ $e/\dot{e}$ ($e$ is the thermal energy, $\dot{e}$ is the
cooling rate) and $t_c << t_h$.  In addition, initial and boundary conditions
can have a strong effect on the simulations.  If these conditions are not
known from observations or if the simulation must be constrained in the choice
of these conditions for computational expediency, the veracity of the results
relative to the actual astrophysical system will be comprised to greater or
lesser degrees.  Thus the simulation studies alone can not be expected to
fully articulate the possible behaviors of astrophysical plasma systems. 

However, direct observation of dynamical plasma astrophysical systems provide
only limited insights into the evolution of the systems because evolutionary
timescales are usually longer than can be observed, or the spatial resolution of the telescope 
does not allow relevant flow features to be clearly identified and tracked. 
Thus, observers must usually be content with inferring the global properties of the flow 
by using snapshots of, for instance, a gravitationally collapsing cloud from
Doppler-shift measurements.  In cases where proper motions can be resolved,
the timescale between observations is usually only long enough to reveal
the first derivative of the motion.  But if a true dynamical
account of the evolution is the goal, then it is the acceleration and structural
changes that matter, as these are linked directly to the forces that drive the flows. 
In addition, observations that measure only proper motions cannot address important
issues such as quantifying the different unstable modes and following those that evolve in a non-linear manner.  
Hence, while proper motion measurements combined with radial velocities define bulk velocities well
within a system, they give effectively no insight into how flows evolve. 

The advent of high-spatial-resolution observing platforms like the HST has, in some
cases, allowed for a more robust observational account of astrophysical plasma
dynamics. For systems that are large enough, close enough, and evolve on
relatively short timescales, telescopes like the HST where the spatial resolution
is $\lesssim$ 0.1$^{\prime\prime}$ in the optical (where the key line diagnostics lie)
can provide multi-epoch observations of objects which, when
strung together, form movies of the evolving plasma.  This technique
has been applied to the Crab Nebula \citep{crabref}, and a handful
of jets from young stars \citep[e.g. HH 30; XZ Tau][]{hm07,xztauref}.
In this paper we have greatly extended this methodology
by using new third-epoch observations to provide a detailed analysis of the
radiative hydrodynamical (and possibly magnetohydrodynamical) behavior of three 
well-resolved, representative HH jets.

Five major classes of behavior are implied by the time-series data:

\begin{itemize}
\item Deflection Shocks, Cavity Evacuation, and Entrainment: Our analysis of
the time-sequence of images in different emission lines shows significant shear
motions and shocks in regions where jets interact directly with
their surrounding clouds.  In HH~1, shear
occurs at the head of the jet where the beam appears to graze the edge of 
the ambient cloud. The bow shock flattens in this interaction region as the head of the
jet deflects and pulls away from the impact point. New knots that formed in the wake
of the deflection shock may represent material in
the process of being entrained by the flow. 
In HH~47, a quasi-stationary ``deflection shock" exists near the base of the
flow where the jet collides obliquely with the edge of an evacuated cavity. This type of
interaction must be common in stellar jets, where ejection angle variations within a given jet
of several degrees are the norm. Additional evidence for prompt entrainment in all three regions 
comes in the form of reverse bow shocks that suddenly ``light up" as slower material is
overtaken by faster jet gas. Knots in the HH~34 jet exhibit spur shocks that accelerate material
at the edge of the jet beam, and excitation maps in the HH~1 jet reveal a similar behavior.

\item Intersecting Shock Waves, Mach Stems, Clumps and Sheets:  In a number of
cases we observe complex shock patterns as they evolve in the flow, though when interpreting
these changes we must keep in mind that the emission is optically thin, so that line of
sight effects come into play.  In HH~2, 
bow shocks ``drape" over each of the multiple fast moving clumps, and the clumps
appear to be in the process of merging into a larger sheet that may form a new, large
bow shock.  The large bow shocks of HH~1, HH~34, and HH~47 contain intricate
networks of fine structure in the form of knots and filaments. The multiple sheets
in the HH~34 bow shock are particularly intriguing, as bright areas with distinct
cooling zones form where the sheets intersect. These knots exhibit anomalous ``pattern''
motions as the intersection points move, and could be the first astrophysical example of
Mach stems. In the HH~47 bow shock, what initially seemed to be a collection of 
dense bullets now seems to be a network of overlapping shocks that come and go
in a manner that resembles turbulence, and similar morphologies are present in the
HH~1 bow shock. 

\item Mach disk structures: Our HST images show two well-resolved Mach disks,
in HH~47A and in HH~34S, but these two objects exhibit markedly different, and highly variable
morphologies.  The Mach disk in HH~47A generally appears close to the standard picture of
a planar shock embedded within a bow shock, while the Mach ``disk" in HH~34S is more
like the point of a reverse bow shock. In 3-D numerical simulations,
the geometry of the working surface often consists of a mixture of several oblique
and normal shocks that may fragment to produce clumpy structures and filaments.
Simulations also show that the Mach disk region can be highly variable in
time, and that working surfaces may collapse into plugs when the gas cools
strongly.  Our work validates and challenges both these kinds of studies by
providing new modes of behavior which must now be addressed with simulations.
 
\item Formation of New Jet Knots Near Source: The images show that the
inherent clumpy nature of the jet begins near the jet source, and each
jet continues to eject new knots.  We can trace the knots closest to the
source in HH~34, and each of the three epochs has a strong knot located
at about 100~AU from the source. If this knot is indeed quasistationary
(like the `sprite' in the Crab nebula), it would support the idea that jets
generate plasmoids by magnetic reconnection processes. Further study with improved
time coverage is the only way to address this issue.

\item Emission and Cooling Dynamics: In all the items above
we are able to track the emission structure in H$\alpha$ and [S~II] of the complex jet flows in
considerable detail, both spatially and temporally.  We have found
several examples of knots that fade when material is no longer fed into the shocks.
The timescale for fading is about a decade, consistent with models of cooling zones
behind radiative shocks. We also observe new shocks as they form. These knots
immediately create strong H$\alpha$ as neutral material enters the shock, and then
later develop cooling zones of [S~II] emission. An extended region of ionized gas
ahead of the strongest shocks in HH~2 appears to be the first clear example
of a radiative precursor in stellar jets.

\end{itemize}

Taken together, our results paint a picture of jets as remarkably heterogeneous
structures that undergo highly-structured interactions between material
within the outflow and between the jet and the ambient gas.  This contrasts with
the bulk of existing simulations which depict jets as smooth plasma systems. 
Recent simulation studies have begun probing how jets evolve as
inherently clumpy systems i.e. as "hypersonic buckshot" \citep[e.g.][]{yirak08,yirak09,yirak10},
but our results show that much more work needs to be done in this domain. 
The structure of jets is a question of fundamental importance because it
speaks directly to processes occurring close to the star-disk system where the
jet is launched, and considerable uncertainty remains as to how this process operates.
Thus, understanding when jet heterogeneity, visible in observations such as
the ones we present in this paper, is first imposed (at launch or via
instabilities above the launch region) is a question our results can address
when combined with simulations.  Beyond quite generic issues of jet
launching (given the ubiquity of jets in astrophysics) our results bear on
equally ubiquitous issues of shock-clump interactions, radiative cooling zones, 
radiative precursors, and shock-shock interactions in the form of Mach stems created at
shock intersection surfaces.  These are issues which can be expected to arise in any
heterogeneous environment with imposed hypersonic flows.  Supernovae, supernova
remnants, molecular cloud dynamics, luminous blue variables, planetary nebulae,
and the environments associated with both the broad and narrow lines regions
of active galactic nuclei all fall into this category.

Thus, the results provided in this work contain information that can, and
should be, used to constrain not only simulations of jet behavior but also
more general studies of astrophysical plasma dynamics associated with shocked
heterogeneous flows. The challenge for modelers will be to separate out the 
``climate" from the ``weather". The goal must be to articulate general principles
that can be learned through this dataset rather than simply attempting to
provide a match to an arbitrary set of initial conditions.

We note in closing that High Energy Density Laboratory Astrophysics (HEDLA)
experiments are extremely relevant to the questions our study raises.  Recent
experiments that track the evolution of multiple clumps with a global shock
whose radius is much larger than the radius of the clumps reveal morphologies
with intriguing similarities to what is seen in our observations \citep{douglas08}.  In
particular, the development of merged "draping" bow-shocks and Mach stems
appear to form in the experiments with patterns that indicate similar
processes as the astrophysical environments.  Future work will focus
specifically on applying the results of these experiments to the behaviors seen in the
clumpy flows presented in the current work.

\acknowledgements
We thank the STScI TAC for granting observing time for this project, and Jay Anderson for
providing his distortion-correction codes. This work is
based on observations with the NASA/ESA Hubble Space Telescope obtained at the Space
Telescope Science Institute, which is operated by the Association of Universities
for Research in Astronomy, Incorporated, under NASA contract NAS5-26555.
Support was provided by NASA through a grant from the Space Telescope
Science Institute, which is operated by the Association of Universities for
Research in Astronomy, Incorporated, under NASA contract NAS5-26555, and by DOE
through the NNSA-sponsored NLUF program for University research.

\begin{center}
\begin{deluxetable}{lcccc}
\singlespace
\tablenum{1}
\tablewidth{0pt}
\tablecolumns{5}
\tabcolsep = 0.08in
\parindent=0em
\tablecaption{}
\tablehead{
\colhead{Object} & \colhead{UT Date} & \colhead{Filter} & \colhead{Exposure Time (ks)} & 
\colhead{HST Program}
}
\startdata
HH 1 \& 2 & 1994.61 & 656N & 3.0  & GO-5114\\
        &         & 673N & 3.0  &        \\
        & 1997.58 & 656N & 2.0  & GO-6794\\
        &         & 673N & 2.2  &        \\
        & 2007.63 & 656N & 2.0  & GO-11179\\
        &         & 673N & 1.8  &        \\
HH 34   & 1994.71 & 656N & 3.0 & GO-5114\\
        &         & 673N & 3.0 &        \\
        & 1998.71 & 656N & 13.4& GO-6794\\
        &         & 673N & 13.4&        \\
        & 2007.83 & 656N & 9.6 & GO-11179\\
        &         & 673N & 9.6 &        \\
HH 47   & 1994.24 & 656N & 11.9 & GO-5504\\
        &         & 673N & 11.9 &        \\
        & 1999.16 & 656N & 4.1 & GO-6794\\
        &         & 673N & 4.1 &        \\
        & 2008.00 & 656N & 7.5 & GO-11179\\
        &         & 673N & 7.5 &         \\
\enddata
\end{deluxetable}
\end{center}
\null\vfill\eject

\begin{figure} 
\def\thefigure{1}
\includegraphics{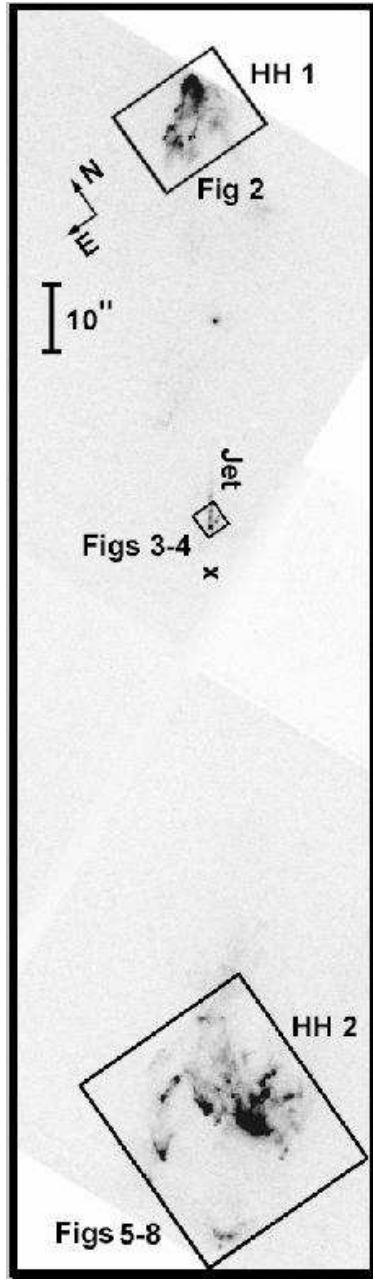}
\caption{Combined [S~II] 673N + H$\alpha$ 656N HST image HH 1 \& HH 2 for the third epoch.
Up corresponds to a position angle of 325 degrees.  The main features are marked,
as are the regions magnified in subsequent figures.
The embedded source that drives the system is marked with an `x' \citep{chini01}.
} 
\end{figure}
\null\vfill\eject

\begin{figure} 
\def\thefigure{2}
\includegraphics{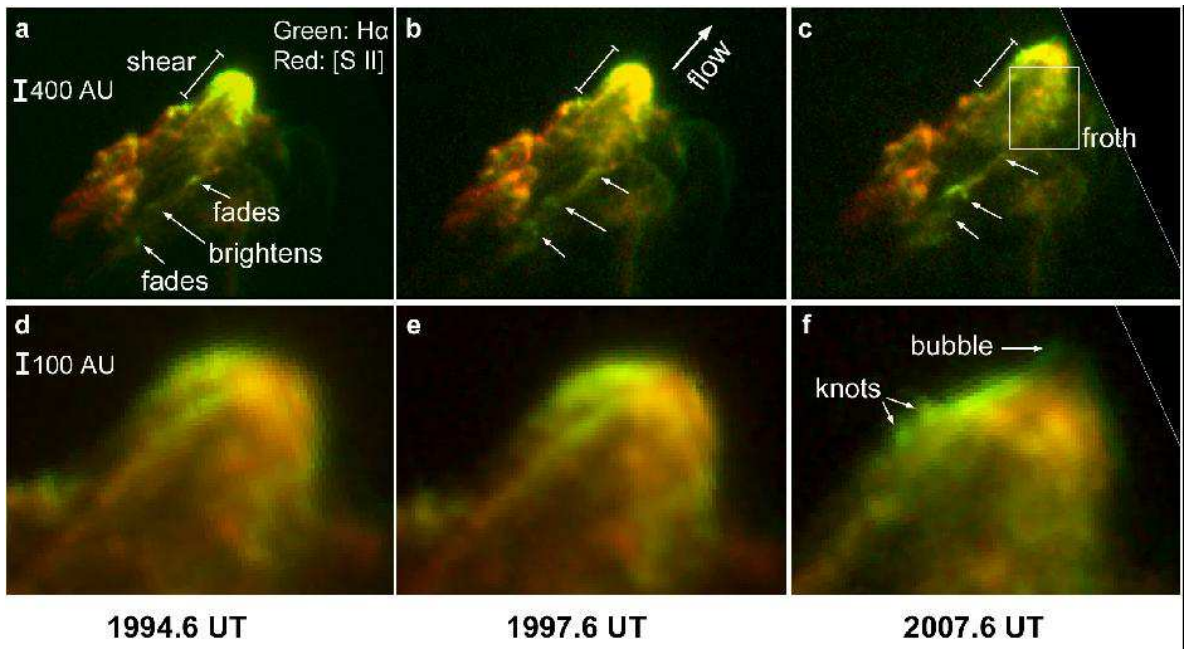}
\caption{Three epochs of HST images of HH 1. The H$\alpha$ image is green, the [S II] image
is red, and yellow denotes emission in both filters. The scale bar assumes a distance of
414~pc. North is up and east to the left.  The top three frames are aligned spatially,
and show that the system moves to the upper right (northwest). The angled scale bar has a fixed length,
and demonstrates that the leading bow shock pulls away from material on its left shoulder. Several
knots along the jet's axis marked with arrows brighten and fade, mostly in H$\alpha$. Changes in
the structure of the bow shock in the third epoch, and the complex structure in the region
denoted `froth' are discussed in the text. The narrow angled white line at the upper right corner
of frames c and f mark the edge of the CCD. The direction of the flow in this, and subsequent figures,
is shown with an arrow.
}
\end{figure}
\null\vfill\eject

\begin{figure} 
\def\thefigure{3}
\includegraphics{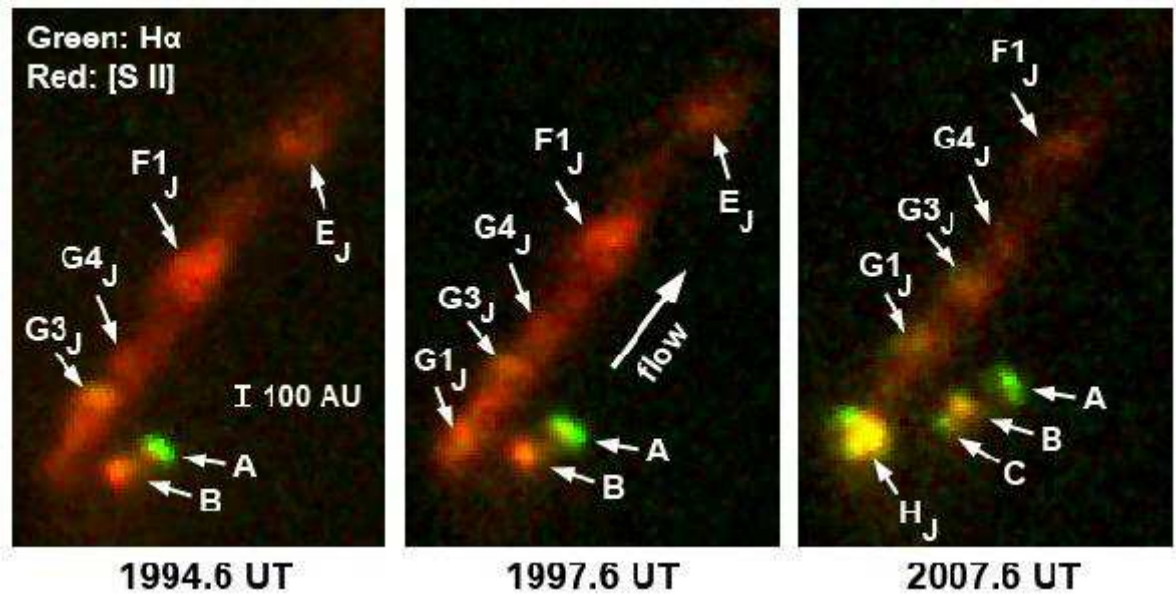}
\caption{Same as Fig. 1 but for the jets in HH 1. North is up and east to the left.
The nomenclature follows that of \citet{hh1pmot}.
Knots with a subscript J are part of the HH 1 jet, and knots without subscripts belong to the HH 501
jet.  Knot H$_J$ in HH 1 and knot C in HH 501 have emerged since the second epoch. The image for the
third epoch is scaled differently from the first two in order to show the structure within knot H$_J$.
}
\end{figure}

\begin{figure} 
\includegraphics{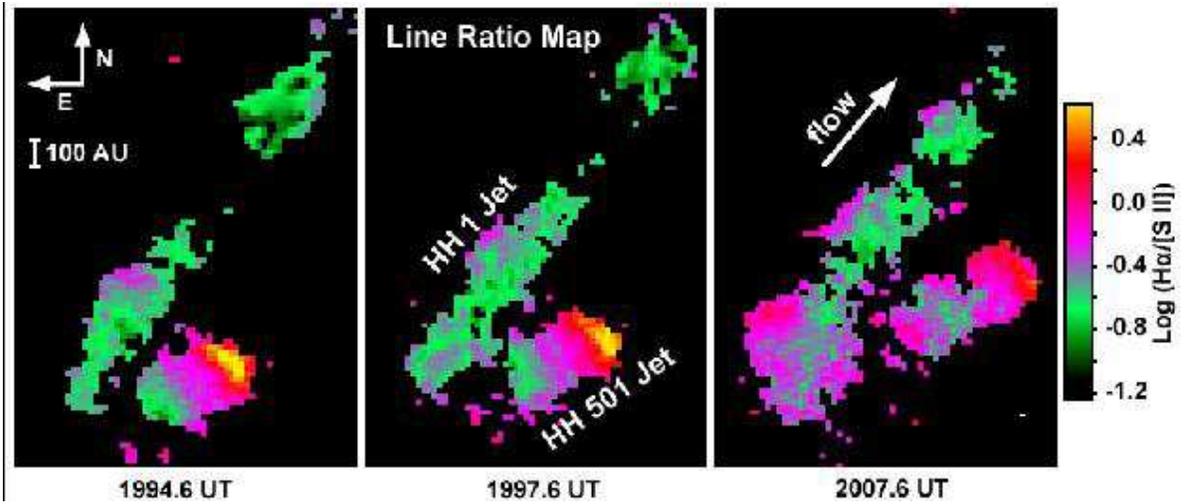}
\caption{Spatial map of line ratios in the HH 1 and HH 501 jets (cf. Fig.~3), for all pixels having
at least 5 ADU above background. The ratio derives from
counts above background in the 656N and 673N filters, corrected for the filter responses.
}
\end{figure}
\null\vfill\eject

\begin{figure} 
\def\thefigure{5}
\begin{center}
\includegraphics{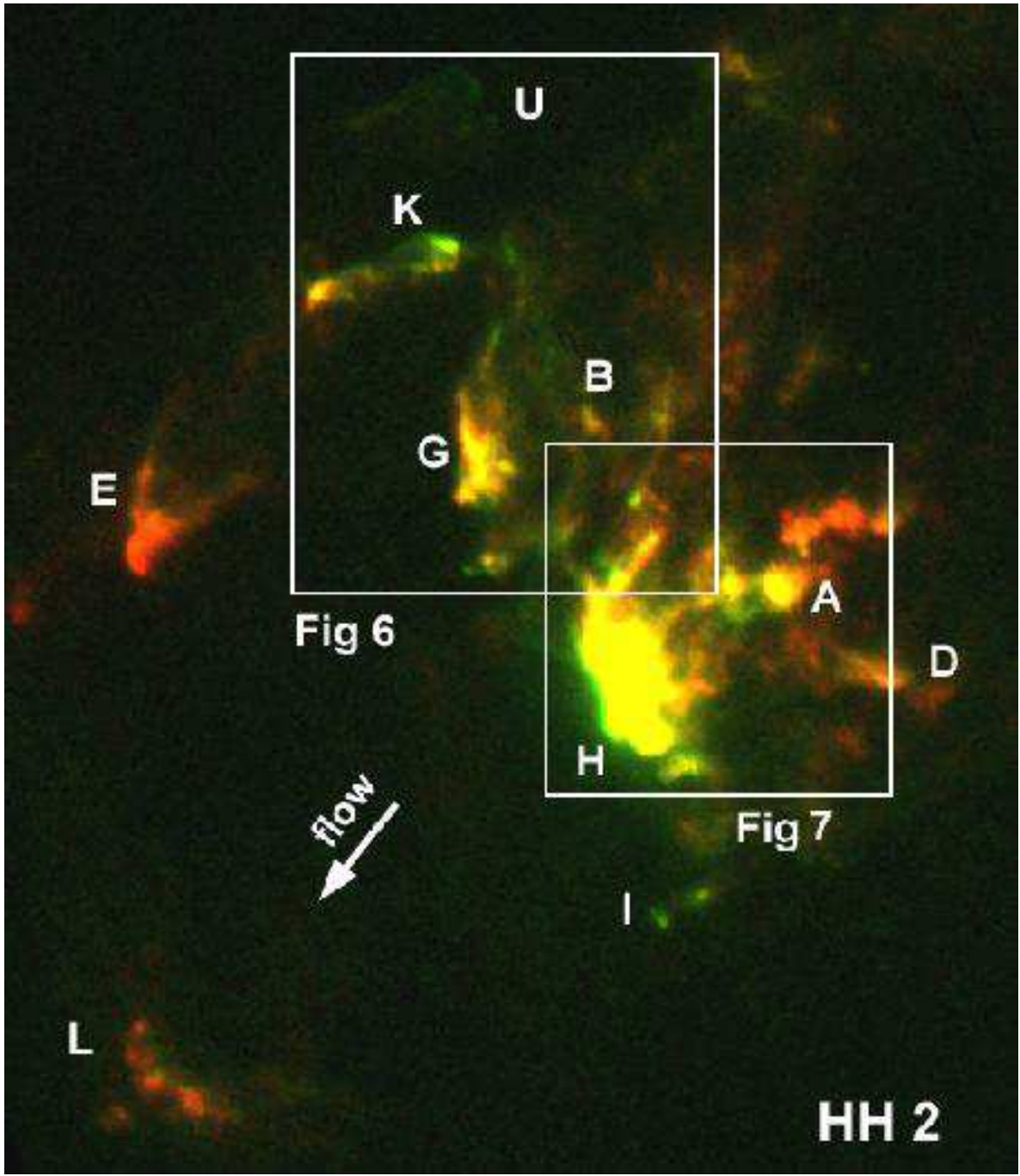}
\end{center}
\caption{Composite H$\alpha$ + [S~II] image for the 2007.6 UT epoch image of HH 2. North is up and east to the left.
The main knots identified by \citet{hh1pmot} are shown, as are the boundaries of subsequent figures of the
region.
} 
\end{figure}
\null\vfill\eject

\begin{figure} 
\def\thefigure{6}
\includegraphics{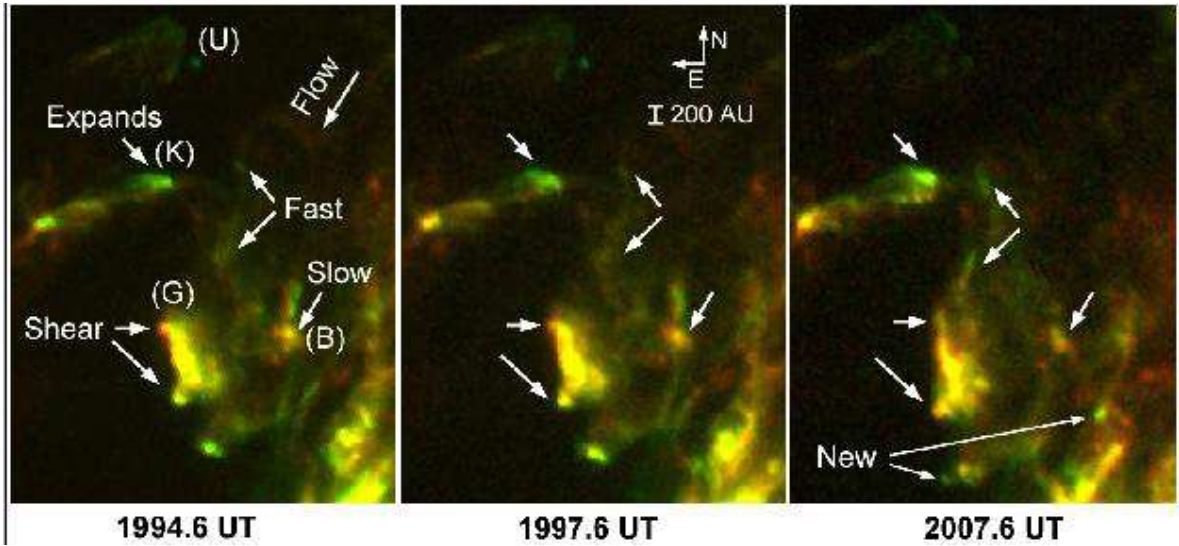}
\caption{Expansion of a portion of Fig. 5, covering the northern portion of HH 2, for
the three epochs.  The main structural variations in the line emission are marked, and are
discussed in the text.
} 
\end{figure}
\null\vfill\eject

\begin{figure} 
\def\thefigure{7}
\includegraphics{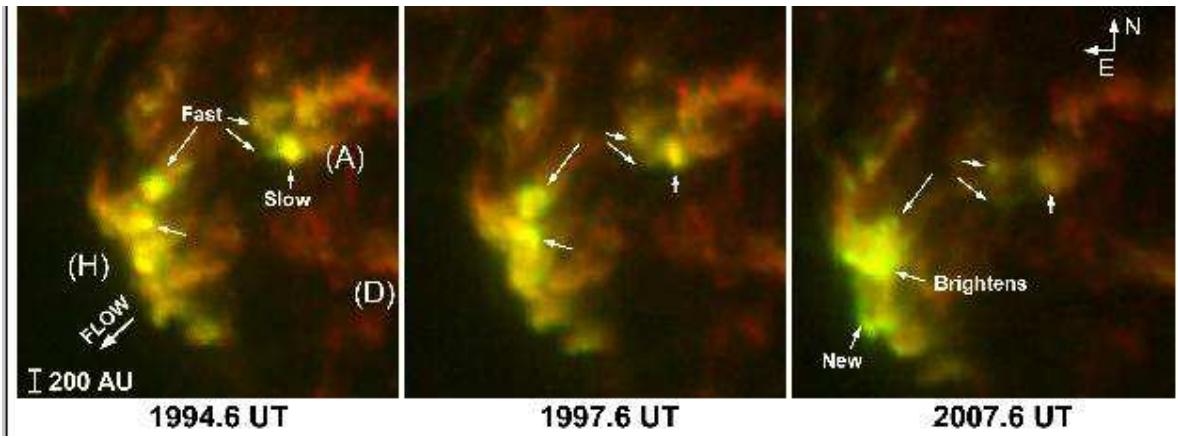}
\caption{Expansion of a portion of Fig. 5, covering knots A and H in the southern portion of HH 2.
The bright working surface HH 2H to the left of the figures consists of numerous clumps
and small bow shocks, while knot A to the right splits into a fast and a slow component. A portion
of knot H located where the wings of two small bow shocks intersect has brightened considerably in
the third epoch, and may represent a Mach stem (see text).
 } 
\end{figure}
\null\vfill\eject

\begin{figure} 
\def\thefigure{8}
\includegraphics{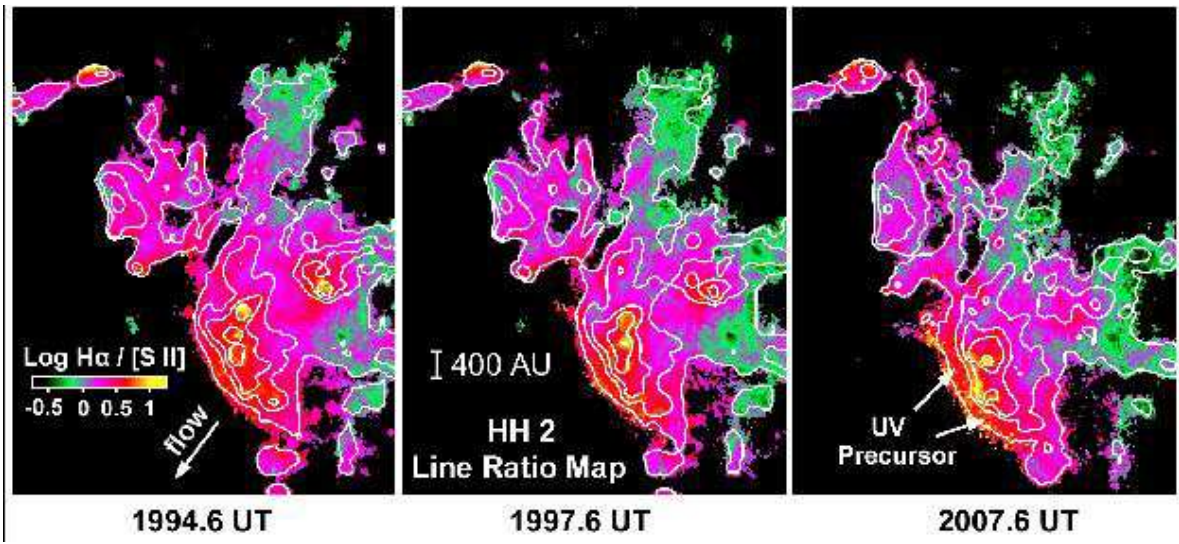}
\caption{H$\alpha$ / [S~II] line ratios within HH~2 (see Fig.~5), plotted for all points where the sum of
the background-subtracted fluxes in the two filters exceeds 20 ADU. The white contours are for the
H$\alpha$ + [S~II] image for each epoch, with the lowest contour equal to 30 ADU, and each higher 
contour a factor of four larger than the previous one. The line ratios tend to increase towards the
leading shock fronts at left. An extended zone of diffuse H$\alpha$ emission (lowest contour)
lies upstream of the brightest knots, most likely arising from the ultraviolet radiation emitted as 
the postshock gas cools.
} 
\end{figure}
\null\vfill\eject

\begin{figure} 
\def\thefigure{9}
\includegraphics{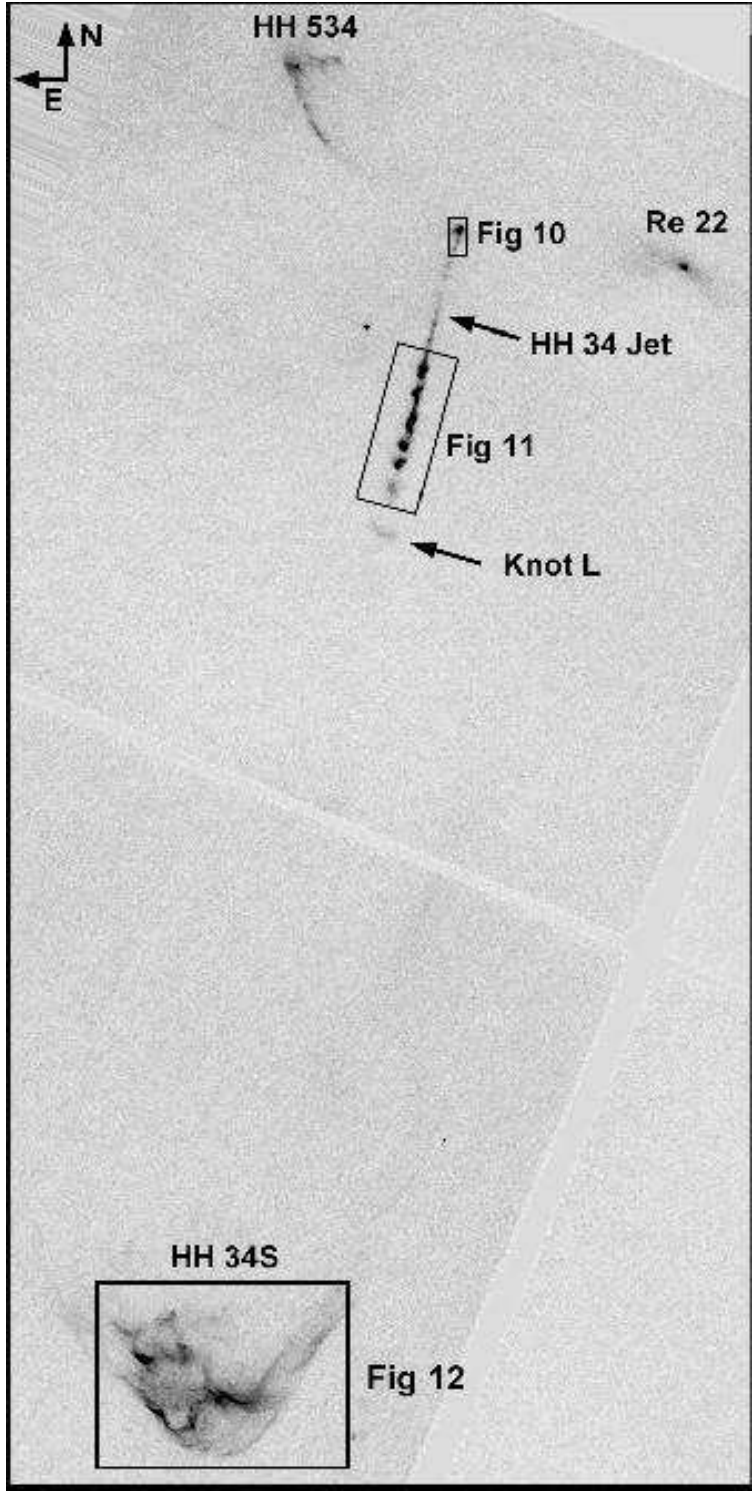}
\caption{Combined H$\alpha$ + [S~II] image of the HH~34 region taken 2007.8 UT. The principal objects
and areas of subsequent figures are marked.
} 
\end{figure}
\null\vfill\eject

\begin{figure} 
\def\thefigure{10}
\includegraphics{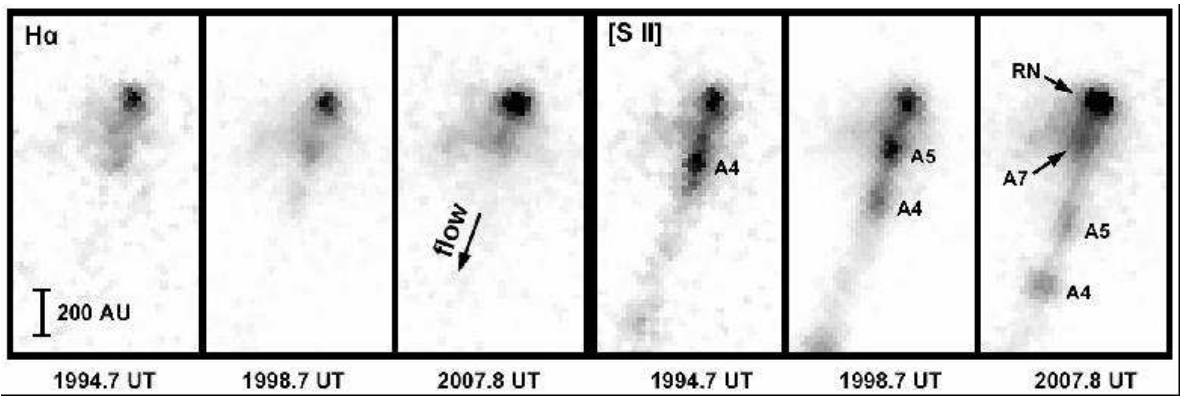}
\caption{H$\alpha$ (left three panels) and [S II] (right three panels) images of the source
region of the HH~34 jet. The object labeled `RN' is a reflection nebula located near the 
exciting source. Labeled features are discussed in the text. North is up and east to the left.
The length of the scale bar assumes a distance of 414~pc.
} 
\end{figure}
\null\vfill\eject

\begin{figure} 
\def\thefigure{11}
\includegraphics{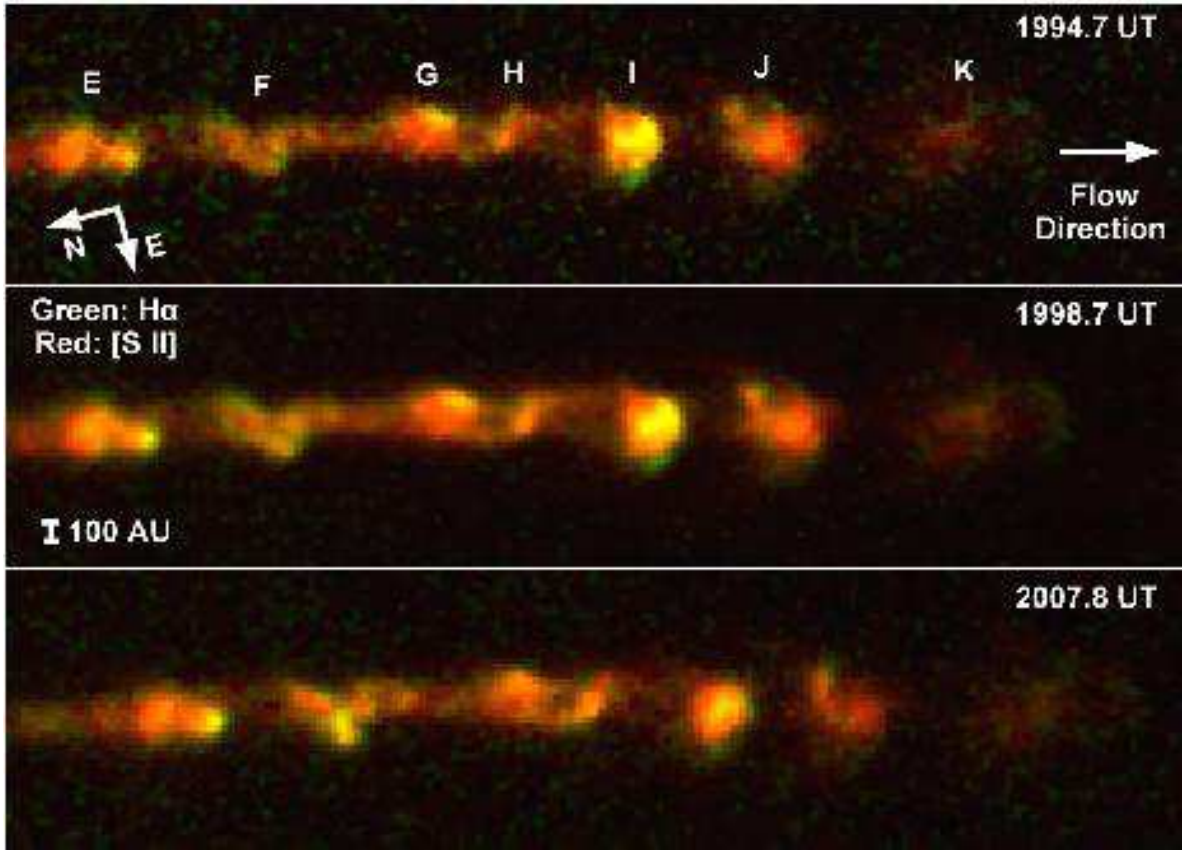}
\caption{Color composite of the brightest portion of the HH 34 jet. H$\alpha$ is in green, [S~II] in
red, and yellow denotes emission in both filters. The flow moves from left to right. The bottom
of knot F has become nearly pinched off in the last image. Knots I, J, and K all faded
significantly between 1998 and 2007.
} 
\end{figure}
\null\vfill\eject

\null\vfill
\begin{figure} 
\def\thefigure{12}
\includegraphics{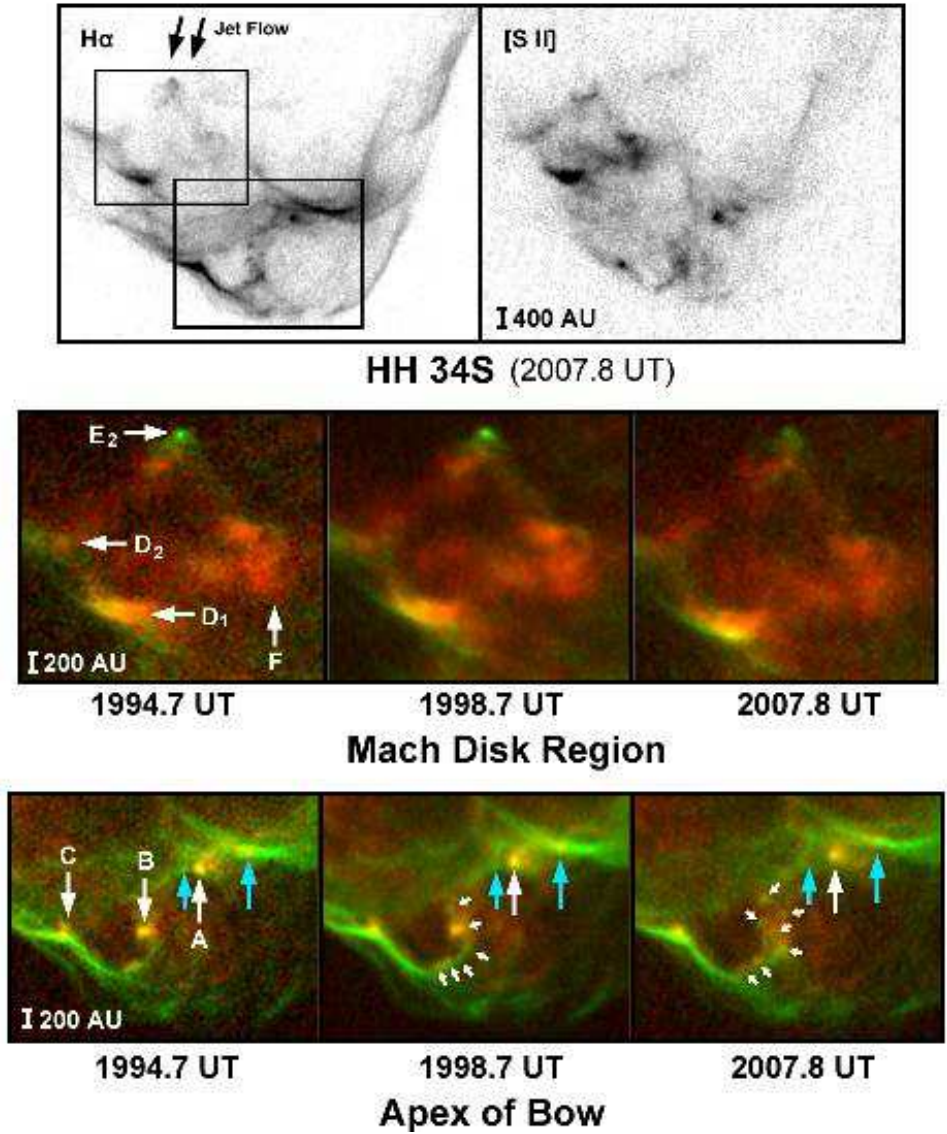}
\caption{Motions within the HH 34S bow shock. Top: Third-epoch
H$\alpha$ and [S~II] images of HH 34S. Arrows mark the projected location where
the jet impacts the bow shock. The two boxes are expanded in the middle and bottom
panels. Middle: The Mach disk region of HH~34S. Green is H$\alpha$ and red is [S~II].
Knot E$_2$ is the Mach disk. Feature D$_1$, located behind the intersection point
of two bow shocks, brightened in the third epoch. Bottom: Same as middle but for the apex
of the bow shock. Knot A, marked with a white arrow,
exhibits lateral motion relative to other nearby features such as the two marked
with blue arrows. Several transient knots, identified with small arrows,
appear near feature B. Knot C lies just behind the intersection of two arc-shaped shocks.
North is up and east to the left in all images.
} 
\end{figure}
\null\vfill\eject

\begin{figure} 
\def\thefigure{13}
\begin{center}
\includegraphics{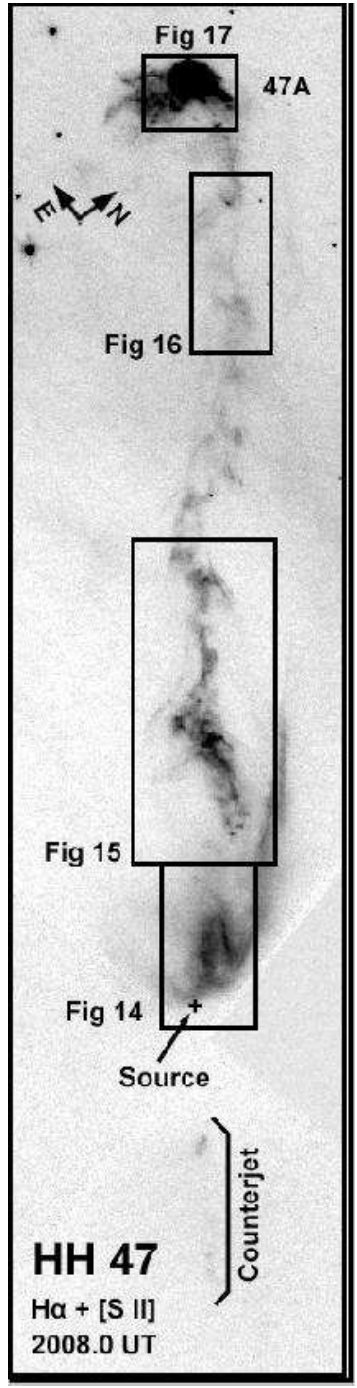}
\end{center}
\caption{Composite H$\alpha$ + [S II] third-epoch image of the HH 47 jet displayed on a logarithmic
intensity scale showing the location of the source and the bright bow shock HH 47A.
Boundaries of subsequent figures are shown.  The PA of the jet is 53.34 degrees.
} 
\end{figure}
\null\vfill\eject

\begin{figure} 
\def\thefigure{14}
\includegraphics{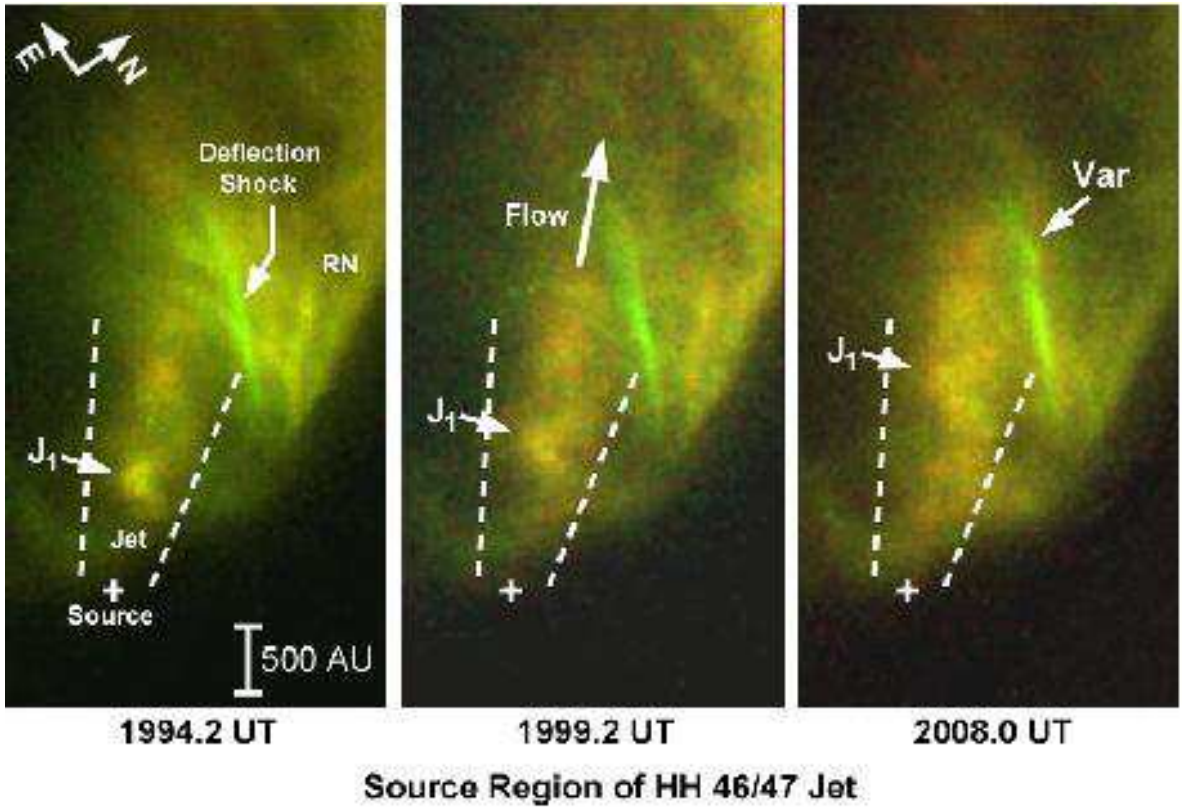}
\caption{The source region of the HH 46/47 jet. H$\alpha$ is green and [S~II] red.
The orientation of the images is the same as the large scale of the jet, PA = 54.3 degrees.
The label RN marks the location of an extended reflection nebula,
a `+' shows the position of the infrared source,
and dashed lines show the approximate extent of the jet. The deflection shock is stationary
over the period, while knot J$_1$ moves along the jet and develops a sawtooth morphology.
} 
\end{figure}
\null\vfill\eject

\begin{figure} 
\def\thefigure{15}
\includegraphics{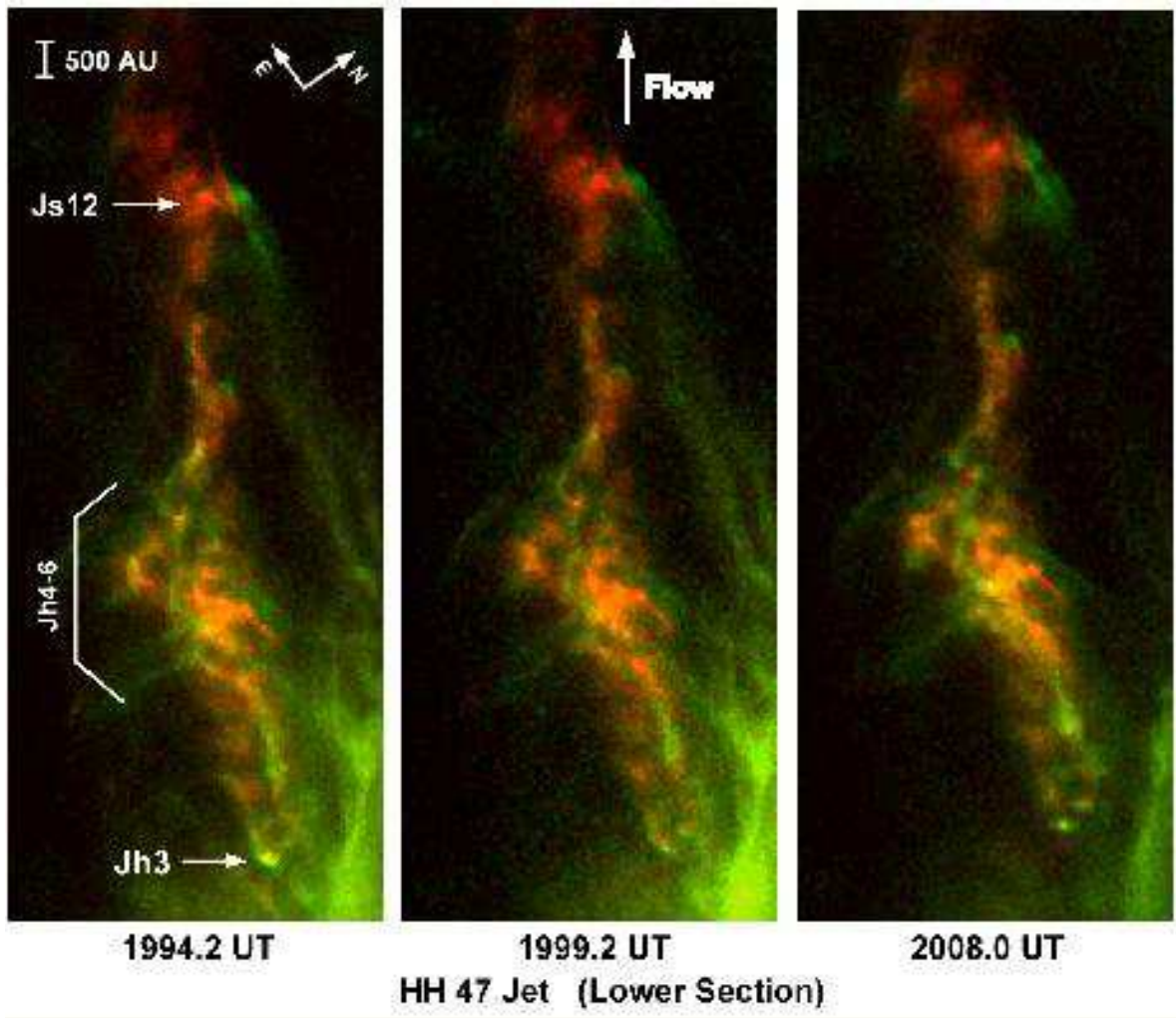}
\caption{Same as Fig. 14 for the lower section of the HH 47 jet. 
Regions identified in the figure are discussed in the text.
} 
\end{figure}
\null\vfill\eject

\begin{figure} 
\def\thefigure{16}
\null\includegraphics{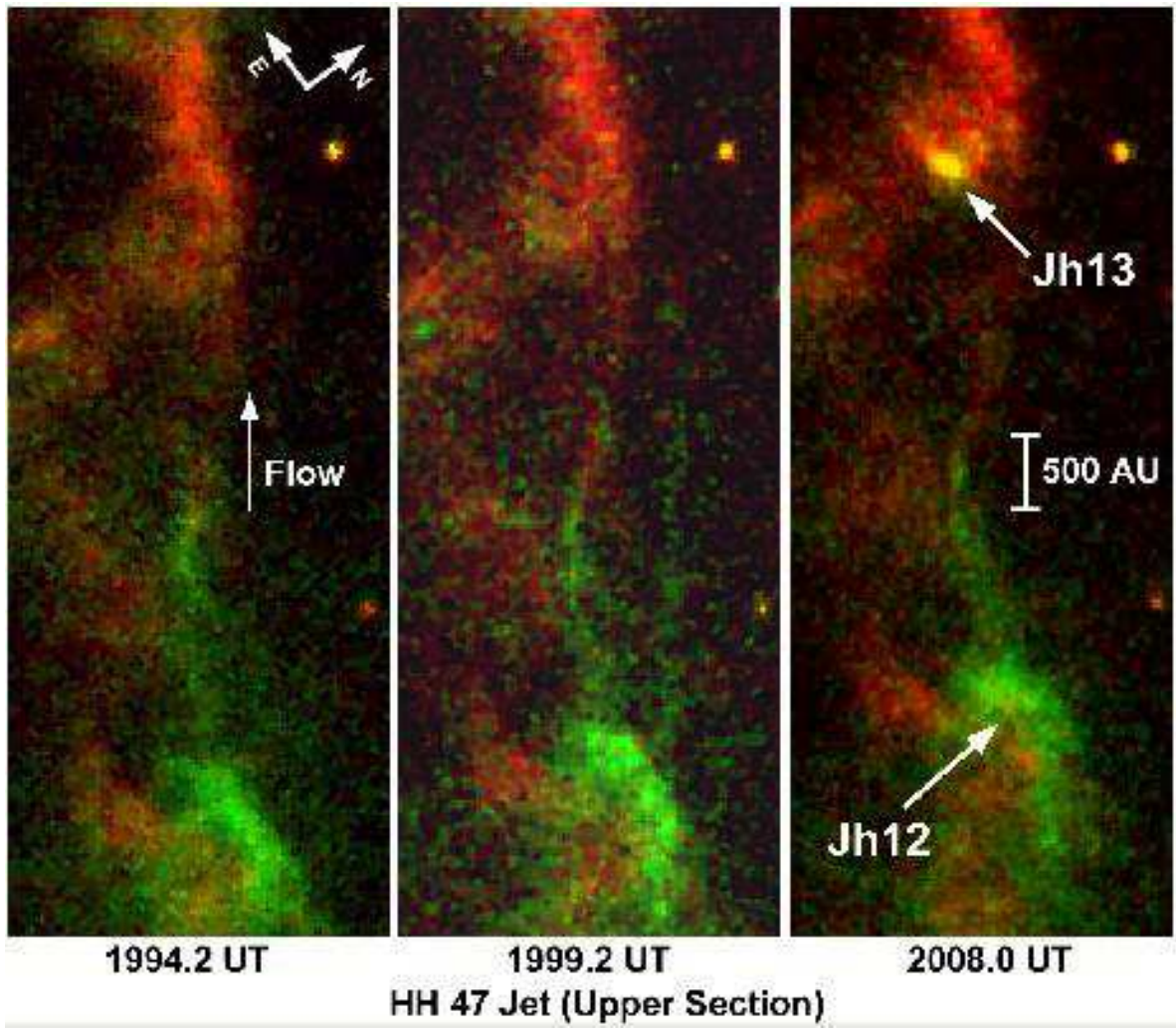}
\caption{Same as Fig. 14 for a portion of the upper section of the HH 47 jet. A new knot, Jh13,
appeared in the third epoch.
} 
\end{figure}
\null\vfill\eject

\begin{figure} 
\def\thefigure{17}
\null\includegraphics{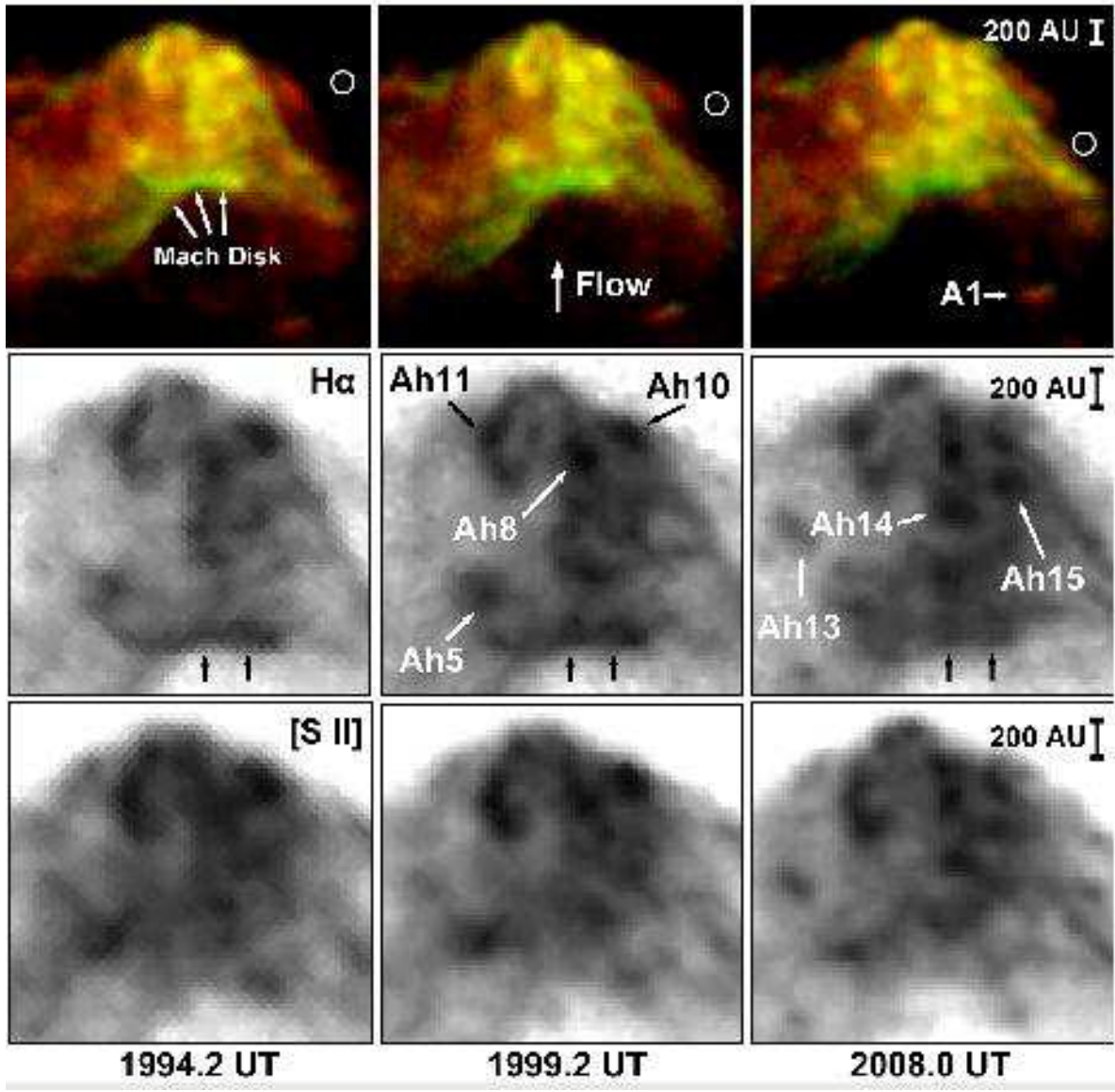}
\null\vskip 5.5in
\caption{The bow shock and Mach disk HH~47A. The bow shock travels from the bottom to
the top in all panels, which have been registered to align the 
Mach disk near the center.  The circled object is a star.
Top: Color composite, with H$\alpha$ in green
and [S~II] in red for the three epochs. 
Middle and Bottom: Logarithmic display of H$\alpha$ and [S~II] images, respectively,
of the central portion of HH~47A.  The structure of the bow shock
is filamentary and highly time-variable.  Features with arrows and labels are
discussed in the text.
} 
\end{figure}
\null\vfill\eject

\end{document}